\pdfoutput=1
\documentclass[a4paper,fleqn,usenatbib]{mnras}
\usepackage{newtxtext,newtxmath}

\usepackage[T1]{fontenc}
\usepackage{ae,aecompl}

\usepackage{graphicx}	
\usepackage{amsmath}	
\usepackage{amssymb}	

\def\ns{neutron star}
\def\wdf{white dwarf}

\title[Predicting Lensing Events by Stellar Remnants]{Predicting Gravitational Lensing by Stellar Remnants}

\author[A. J. Harding et al.]{
Alexander J. Harding,$^{1,2}$\thanks{E-mail: Alexjharding6@gmail.com}
R. Di Stefano,$^{1}$
S. L\'epine,$^{3}$
J. Urama,$^{4}$
\newauthor
D. Pham$^{5}$ 
and C. Baker$^{1,2}$
\\
$^{1}$Harvard-Smithsonian Center for Astrophysics, 60 Garden
  Street, Cambridge, MA 02138\\
$^{2}$School of Physics \& Astronomy, University of Southampton,
  Southampton, SO17 1BJ, UK\\
$^{3}$Department of Physics \& Astronomy, Georgia State
  University, 25 Park Place, Atlanta, GA 30302\\
$^{4}$Department of Physics \& Astronomy, University of Nigeria,
Nsukka, Enugu State 41001\\
  $^{5}$Cornell University, Ithaca, NY 14850\\
}

\date{Accepted XXX. Received YYY; in original form ZZZ}

\pubyear{2017}

\begin{document}
\label{firstpage}
\pagerange{\pageref{firstpage}--\pageref{lastpage}}
\maketitle

\begin{abstract}
Gravitational lensing provides a means to measure mass that does not rely on detecting and analysing light from the lens itself. Compact objects are ideal gravitational lenses, because they have relatively large masses and are dim. In this paper we describe the prospects for predicting lensing events generated by the local population of compact objects, consisting of $250$ neutron stars, $5$ black holes, and $\approx35,000$ white dwarfs. By focusing on a population of nearby compact objects with measured proper motions and known distances from us, we can measure their masses by studying the characteristics of any lensing event they generate. Here we concentrate on shifts in the position of a background source due to lensing by a foreground compact object. With \textit{HST}, \textit{JWST}, and \textit{Gaia}, measurable centroid shifts caused by lensing are relatively frequent occurrences. We find that $30-50$ detectable events per decade are expected for white dwarfs. Because relatively few neutron stars and black holes have measured distances and proper motions, it is more difficult to compute realistic rates for them. However, we show that at least one isolated neutron star has likely produced detectable events during the past several decades. This work is particularly relevant to the upcoming data releases by the \textit{Gaia} mission and also to data that will be collected by \textit{JWST}. Monitoring predicted microlensing events will not only help to determine the masses of compact objects, but will also potentially discover dim companions to these stellar remnants, including orbiting exoplanets.

\end{abstract}




\section{Introduction}

Black holes, neutron stars, and white dwarfs are the end states of
stellar evolution. With few mass measurements of isolated compact
objects, we do not yet know the distributions of their initial
masses. Furthermore, mass transfer from a companion can produce mass
increases, but we have only a partial understanding
of the relevant circumstances.
Some compact objects eventually participate
in mergers, potentially producing gravitational waves \citep{PhysRevLett.116.061102}, short gamma-ray
bursts, and Type Ia supernovae. There are therefore many reasons to
measure the masses of compact objects. Because most are dim across
multiple wavebands, it has been difficult to discover more than a
small fraction of them, let alone to measure their masses.

As it happens, mass measurement remains one of the most challenging
tasks in observational astronomy. One generally relies on the detected
orbital motion of a stellar or planetary companion, or on the presence
of an orbiting disk of gas, dust, or debris. 
Emission lines emanating
from a compact object can also be used to estimate its mass, but this
has been done in only a relatively small number of cases 
\citep{0004-637X-757-2-116}. 
Gravitational lensing provides an alternative approach to
mass measurement. It has the advantage of only relying on the light
from a background source, and can therefore be employed even for dark
lenses. In fact, since light from the lens can interfere with the
detection of lensing effects, compact objects are ideal lenses. Of the
roughly 18,000 unique lensing events detected to date, it is expected
that a few per cent must have been caused by black holes or neutron
stars, and about $10-15$ per cent were likely caused by white dwarfs
\citep{0004-637X-684-1-59}.
The complication is, however, that we do not know
which of the detected events involved compact lenses.

We propose to circumvent this problem by identifying local compact
objects and predicting when they are going to produce lensing events,
so that these events may be studied as they occur. By focusing on
pre-selected compact objects in the near vicinity of the Sun, we ensure
that the lensing event will be caused by a white dwarf, neutron star,
or black hole. Furthermore, the distance and proper motion of the lens
can be accurately measured prior to the event, or else
afterwards. Armed with this information, the lensing light curve
allows one to accurately measure the mass of the lens; this is because
the Einstein radius crossing time is determined from the model fit to
the lensing event, and when the proper motions and distances to the
source and lens are known, its value depends solely on the mass of the
lens.

In addition, deviations to the lensing light curve can reveal dark
companions orbiting the lens, including planets. A small number of millisecond
pulsars are known to have planets 
\citep{1992Natur.355..145W,1993Natur.365..817B}, but whether
neutron stars or black holes generally host planets remains an open
question, and the circumstances under which these planets might form
remain unclear. The microlensing discovery of planets orbiting neutron
stars or even black holes would provide critical data for addressing
these problems. White dwarfs, on the other hand, are expected to host
retinues of surviving planets. Indeed, white dwarfs are known to
interact with debris disks or asteroids, and the impact from accreting
asteroids may soon be detectable \citep{0004-637X-712-1-142}. Yet, planets
orbiting white dwarfs have not yet been discovered. The prediction and
subsequent observation of gravitational lensing events would allow us
to systematically search for planets orbiting compact objects, while
at the same time measuring the masses of both the stellar remnants and
their dim companions.

In the next section, we provide background on gravitational lensing and explain how we can estimate the rate at which stellar remnants can produce these events. In \S 3 we discuss how we obtained our lists of compact objects used in our research, and the source catalogues we have used to obtain the background sources in the lensing candidates' vicinity.
In the following section, \S 4, we present the results of the calculated event rates using the measured background source densities. We then give the process of determining events for indivudal lenses by projecting their motion forward in time.  
\S 5 then explains the method by which we hope to map out predictions of future lensing events. For demonstrative purposes, we give a neutron star example, that has appeared to cause two events in the past, and a white dwarf which may come close to one of two stars in $\sim$ 2040. \S 6 is devoted to a summary of our conclusions.

\section{Gravitational Lensing Events}
\subsection{Einstein angle and required angles of approach}
When there is perfect alignment between a point-like source at distance $D_S$, 
an intervening point-like mass, $M$, at distance $D_L$, and an observer,
the image of the source is a ring whose angular radius, $\theta_E,$ is
referred
to as the Einstein radius 

\begin{equation}
  \theta_E  =
  \sqrt{\bigg(\frac{4GM_{L}}{c^2}\bigg)\bigg(\frac{D_L-D_S}{D_LD_S}\bigg)}
  \\
  \simeq  10~{\rm
    mas}~\sqrt{\bigg(\frac{M_{L}}{1.4~M_\odot}\bigg)\bigg(\frac{0.1~{\rm
        kpc}}{D_L}\bigg)},
  \label{eqn:theta}
\end{equation}
where the expression on the right hand side is valid if $D_L << D_S$. In most cases, the alignment is not perfect, and the angle between source
and lens would be measured by the observer to be $\delta\, \phi$, in the
absence of lensing. Gravitational lensing shifts the source positions and the magnifications of the images. 
The size of these effects depends on the value of $\delta\, \phi$, expressed
in units of $\theta_E$: $u = \delta\, \phi/\theta_E.$   

\subsection{Event Prediction}
When light from a distant star (source) is deflected by an intervening
mass (lens), the image of the source is distorted. The size of the
distortion is typically comparable to or smaller than the Einstein
angle, $\theta_E$, which is on submillisecond scales for most stars in
the Milky Way. Although the resulting {\em astrometric} deflections have
been too small to detect until now, evidence of lensing has been found
through the {\em photometric} effect: the transient brightening of the
background star which occurs when the angular separation between the
source and lens falls below $\theta_E$.

The prediction of a photometric event requires prior knowledge 
of the angular trajectory of the lens relative to the source with submilliarcsec
precision, unless the value of $\theta_E$ is larger than typical. This is indeed 
the case for nearby lenses. As demonstrated below, the value of $\theta_E$
can be several milliarcseconds (mas) for many nearby lenses, and more than $10$ mas for some. For such large Einstein rings, microlensing prediction is much easier. 
The distance of closest approach must be predicted to within about $10$ mas in order
to reliably forecast a lensing event. This is however possible with present 
technology \citep{2012ApJ...749L...6L}. Even in cases without enough data in hand 
to allow a specific prediction of photometric effects, the probability 
of such a close approach can be estimated. If therefore, we identify 
a set of stars with a large combined probability ($P>>1$) of producing a
photometric event within a year, photometric monitoring of this group
will lead to the detection of events caused by one or more of these
known nearby lenses.

For photometric events, we can either predict the time and distances
of closest approach of specific high-probability events, or we can
monitor modest-sized (hundreds to about a thousand) groups of nearby
stars in order to be assured that some of them will act as lenses
during a given year. The advantage of predicting photometric events is
that these can be detected, and their progress tracked, by
ground-based monitoring. Indeed, if the background sources are bright,
even small telescopes can track the progress of the photometric
lensing events. Astrometric events are however much more common and
more easily predicted.

\subsection{Photometric effects}

When both lens and source are point-like, the photometric magnification
is given by:
\begin{equation}
A(u)=\frac{u^2+2}{u\sqrt{u^2+4}}
\end{equation}
where $u$ is the angle between the source and lens, expressed in units
of the Einstein angle, $\theta_E. $ 
For large $u$, $A \sim u^{-4}$ \citep{1986ApJ...304....1P}. 
This means that detectable effects are expected only for approaches within 
a few times $\theta_E.$ The magnification associated with more distant 
approaches can be detected only in exceptional cases, as when there
is high precision monitoring by \textit{Kepler} \citep{0034-4885-79-3-036901} or \textit{TESS} \citep{2014AAS...22411302R}.  

\subsection{Astrometric effects}

The magnitude of the astrometric centroid shift, $\delta(u)$, is given by 
\begin{equation}
\delta(u) = \frac{u}{u^2+2} \theta_E.
\label{eq:centroid1}
\end{equation}

Astrometric effects are potentially detectable for significantly
larger angles of approach between the source and lens 
\citep{2000ApJ...534..213D}. This is because for the astrometric case, 
the magnitude of the centroid shift $\delta(u)$ is proportional to $u^{-1}$, 
whereas for the photometric case, the magnification is given by $u^{-4}$. 

The largest centroid shift occurs for
separations $u=\sqrt{2}$. For large values of $u$,
\begin{equation}
\delta(u) \simeq \frac{\theta_E}{u} =\frac{(\theta_E)^2}{\delta\phi},
\label{centroid2}
\end{equation}
and the magnitude of the astrometric deflection falls off as the
angular separation $\delta\phi$. If $\theta_E$ is $5$~mas, and if the
astrometric centroid of the source can be measured to a precision of $0.1$~mas,
then the approach need only be closer than $250$~mas for an
astrometric deflection to be detected. For somewhat larger Einstein
rings or for more sensitive measurements \citep{Riess:2014uga}, even
comparatively large distances of approach will produce detectable
centroid shifts. This means that events causing detectable centroid
shifts will be much more common than events producing significant
photometric amplifications.

The size of the shift measured at a time when the true separation
between source and lens is known allows one to measure the value of
$\theta_E.$ If we know the distance to the lens, and if we can
estimate the distance to the source (or else assume that $D_L/D_S <<
1$, which is generally the case for nearby lenses), Equation \ref{eqn:theta} can
be used to compute the mass of the
lens.

To measure the
astrometric deflection we must measure the source position multiple times,
for example when
the lens is far away, and then at least once when the lens is near the
distance of closest approach from the source. 
High precision mass measurements require high precision measurements of 
the centroid shift. 

\textit{HST} and \textit{Gaia} are ideal for the measurement of
lens masses.
For \textit{HST} imaging we will assume an 
astrometric precision of $x=0.3$ mas \citep{2011PASP..123..622B}.
We note that significant improvement may be possible; e.g., 
by employing a scanning technique for data acquisition, 
\citet{Riess:2014uga} claim achievable astrometric
precisions of $0.03$~mas with \textit{HST's} WFC3/UVIS camera. 
Consider a neutron star at $100$ pc; $\theta_E \sim 10$~mas.  
If the angular distance of closest
approach is $\approx100$ mas, then the astrometric deflection would be
$\approx1$~mas. This means that \textit{HST} measurements have the potential to make high-precision mass measurements, as can \textit{Gaia}, which is expected to achieve positional precisions
of $10-25$ microarcseconds ($\mu$as) for stars of 13{\it th} 
magnitude, and $\sim 0.25$~mas
for stars of 20{\it th} magnitude.  Thus, the mass of the neutron star
considered above could
potentially be measured to a precision of $0.5-1.6$ per cent.

The relative advantages are that \textit{Gaia} covers the whole sky, and provides
higher precision for bright stars. \textit{HST} only covers a small portion of the sky,
but allows the large population of dim stars to serve as sources whose
deflections can be measured. In addition, \textit{HST} can image dim compact objects,
such as isolated neutron stars, allowing the relative positions of lens
and source to be measured even in these situations.

\subsubsection{Planets} 

Planets orbiting compact objects may also serve as gravitational
lenses. The detectable signatures are largely determined by the
angular separation, $a$, between the compact object and its planet,
expressed in terms of the Einstein angle, $\alpha = a/\theta_E$. One
may identify three modes.

\smallskip

{\bf (a)} Close-orbit planets are in in orbits with $\alpha<0.5$. 
Such planets can produce photometric effects, even when the 
distance of closest approach 
is $\sim 5 \theta_E$, generally too far for the compact object to
produce photometric effects. The lensing signatures  
are quasiperiodic deviations from
baseline \citep{0004-637X-752-2-105}.
Astrometric shifts are also potentially detectable. 

\smallskip

{\bf (b)} The planets that are generally discovered via microlensing 
are located in the so-called `zone for resonant lensing': 
$0.5 < \alpha < 2$. The signatures of planets are
short-lived deviations from the point-source/point-lens light curve 
\citep{1991ApJ...374L..37M,1992ApJ...396..104G}. 
In these cases, the distance of closest approach must be small enough
that the central star also produces detectable effects. 

\smallskip

{\bf (c)} Wide-orbit planets have $\alpha\gtrsim2$. In these cases,
 the signature of
lensing by a planet is similar to lensing by an independent mass
\citep{0004-637X-512-2-579}. The size of the Einstein ring, the event rate per lens,
and the Einstein ring crossing time all scale with $\sqrt{q},$ where
$q$ is the planet-to-lens mass ratio $m_{pl}/M.$ 

It is this last class that has the potential to significantly add to the rate of
lensing events associated with compact objects. First, we expect
such planets to exist:  
planets on distant orbits have a better chance of surviving
the post main-sequence phases, and an individual compact
object may have more than one planet on a wide orbit. 

The rate of short-duration events produced by such wide-orbit planets may be comparable to
(generally several times smaller than) the rate of events caused by the compact
objects themselves. Planetary systems of nearby compact objects may
span several square arcsec. For example, for a neutron star at
$100$~pc, $\theta_E = 10$ mas the Einstein radius is $R_E = \theta_E D_L = 1$~AU; 
thus, any planet at $1000$~AU which acts as an independent 
lens will be $10$ arcsec from the position of the neutron 
star. Planets at these distances could produce either photometric 
or astrometric signatures of lensing. 

Wide-orbit planets can be detected even when the compact object they orbit are not close enough to cause an event themselves \citep{2013ApJ...771...79D}. We do not know the locations of any planets orbiting specific compact
objects. It is therefore impossible to make event predictions 
for specific dates. Nevertheless, it is almost certain that
a significant fraction of compact objects have planets. The positions and magnitudes of stars within a radius equal
to the possible size of the planetary system should be monitored.

\subsubsection{Event Duration} 

The Einstein radius crossing time is
\begin{equation}
\tau_E = \frac{\theta_E}{\mu} = 36.5~{\rm days}
\bigg(\frac{\theta_E}{10~{\rm mas}}\bigg)\bigg(\frac{100~{\rm
    mas~yr^{-1}}}{\mu}\bigg).
\label{eqn:tau}
\end{equation}
Event durations are thus proportional to $\tau_E.$ For photometric
events, detectable deviations from baseline are typically around one
per cent, so that the event duration, $\tau_{dur}$ is a few times
$\tau_E$. For astrometric events, detectable shifts occur for $\theta
< \theta_E/\delta_{min}.$

We define a detectability factor, ${\cal F}$ as
\begin{equation}  
{\cal F} = \frac{\tau_{dur}}{\tau_E}  
\label{eqn:detectabilityfactor}
\end{equation}  
For astrometric events, the value of ${\cal F}$ can be of order $100.$
Some of the white dwarfs in our sample have
$\theta_E\approx30$~mas. If a shift of $0.3$~mas ($0.03$~mas) is
detectable, then the lensed source may be nearly $3$ arcsec
($30$ arcsec) from the white dwarf at closest approach. For
the neutron star Geminga, $\theta_E \approx 6.3$~mas. In order for
astrometric shifts of a background source close to Geminga to be
observable, the lens and source angle of closest approach must
be $< 130$~mas ($<1.3$ arcsec) if $\delta_{min}=0.3$~mas
($\delta_{min}=0.03$~mas). 

\subsection{Event Rates}

The average rate ${\cal R}$ at which an individual mass will lens a
background source can be calculated from: 
\begin{equation}
  {\cal R} ={\cal F} \times 2 ~\theta_E~ \mu~ \sigma,
  \label{eqn:rates}
\end{equation}
where ${\cal F}$ is the detectability factor defined in Equation \ref{eqn:detectabilityfactor},
$\theta_E$ is the Einstein angle for the lens (in arcsec),
$\mu$ is the proper motion of the lens relative to background sources
(in arcsec per year), and $\sigma$ is the average density of
background stars (in stars per square arcsec).
For an individual lens, ${\cal R}$ should be viewed as a long term
average.   
As the lens moves across the sky it crosses
regions with different values of $\sigma$. Furthermore, the value of
${\cal F}$ is larger for bright background stars than for dim
background stars.

\subsection{Types of Prediction}

\subsubsection{Rate computations} 

We can consider an individual mass and compute the rate at which it generates
photometric events and/or astrometric events. 

{\bf Rates based on the background surface density:} 
We employ Equation \ref{eqn:rates} to compute the average
rate at which an individual mass, $M$, produces lensing events. 
The approximate mass of the
lens  and its distance from us, provide enough information to compute the 
area per unit time covered by its Einstein ring and, more generally,
$dA/dt$, the rate at which area is 
covered by $\theta_E \mu$. The surface density of 
background stars, when multiplied by $dA/dt$, provides the rate.
We will consider background stars from catalogues based on the ground-based
surveys: the Sloan Digital Sky Survey (SDSS) and the Two Micron All-Sky Survey (2MASS), as well as on data collected by space missions 
(\textit{Gaia} and \textit{HST}). 

{\bf Rates based on specific approaches:} By keeping track of not only the 
density of stars behind each lens, but also the positions of the stars in the
background, we can compute the distances and times of closest approach
to each background star. By counting the number of approaches per
unit time that lead to detectable events, we are also computing
the average rate. 

Although we typically need additional information to make specific predictions,
rate calculations of both types described above 
can 
identify the compact objects that produce events at the highest rates.
In \S 5 we consider individual lenses with high event rates, and
examine in detail the backgrounds over which they are presently
passing. This process allows us to determine the times of recent or
soon-to-occur events, generated by that particular lens.

\section{Event Rates}

In this section we use Equation \ref{eqn:rates}
to compute average event rates for potential lenses that are
white dwarfs, neutron stars,
and black holes. The specific set of compact objects we consider are only a
small subset of known stellar remnants, because we limit ourselves
to those with measured parallax and proper motion. The parallax provides
the distance $D_L$ to the potential lens. 
For each class of compact object, except the black holes, we use a standard mass value (see \S 3.2).
Assuming that the distant star to be lensed is much farther from us
than the lens, we use Equation \ref{eqn:theta}
to compute the Einstein angle.
The value of $\mu$ is known, and  product $\theta_E \mu$ gives us
 the area per unit time swept
out by  
the Einstein ring of the compact object.  
In \S 3.1 we outline the procedure we used to select the compact
objects.

To estimate the rate, $\mu\, \theta_E \sigma$, at which each compact object can produce lensing events,
we must measure the background surface  density, $\sigma$,
 of stars that could be lensed.
 We have done this
by using four different catalogues: Gaia Data Release 1 (DR1), SDSS, 2MASS,
and the {\sl Hubble Source Catalog} (HSC).
The measurements that produced each catalogue are sensitive to specific
sets of stars. Thus, while some stars are listed in
multiple catalogues, others may appear in just a single catalogue.
In \S 3.2 we describe our use of the catalogues and the derivations
of the surface density of stars associated with each catalogue. 
We assume that the stars in the catalogues are more distant than 
the compact
object that is the putative lens. While some individual catalogued
stars may be closer, most are likely to be farther away. This hypothesis is testable because distance
measurements,  or lower limits on their
distances, can be (or have already been) measured. 

In \S 3.3 we compute the average rates of events, using Equation \ref{eqn:rates}.
Because the stellar densities are different for different types of
background stars,
we compute individual rates for each catalogue. 
We also consider what the rate would be
if we could obtain \textit{HST} images of each field: generally it would be
significantly larger because a large number of dim stars would 
now add to the value of $\sigma.$

Of primary importance is the value of the detectability factor ${\cal F}$, which depends on how the lensing signatures are to be detected.
As discussed above, 
for photometric detection ${\cal F}$ will generally be a factor of a few.
For astrometric detection it is larger and can 
be as large as several hundred. For any individual event, the value of 
${\cal F}$ is influenced by specific circumstances, such as the relative
brightness of the lens and background source.  
In \S 3.4 we present computed values of the event rates, choosing a specific 
astrometric criterion for detectability: a shift of $0.3$~mas is detectable.
Under certain circumstances, smaller shifts can be measured. Ultimately
for example, data from \textit{Gaia} will measure positions for the {\sl dimmest} 
stars with similar precision, and will achieve $5$ $\mu$as precision for 
bright stars. Deviations associated with lensing might have to be
somewhat larger than the precision limit to permit reliable measurements
of the Einstein angle; nevertheless lensing-induced
centroid deviations below $0.3$~mas 
will almost certainly be measurable. 
It is also worth noting that it
 has been suggested that under certain circumstances, the astrometric precision afforded by \textit{HST} 
can also be improved by modifying the observing strategy \citep{Riess:2014uga}.

If there is significant light from the lens, it may not be possible
to take advantage of
high-precision capabilities of \textit{HST} or other high-angular-resolution telescopes. 
\textit{JWST} will not be sensitive to 
blue light from white dwarfs, making it ideal for detecting astrometric
shifts induced by white dwarf lenses.     
Thus, large values of the detectability factor, similar to (and even larger than)
those
shown in Tables \ref{wd-examples}, \ref{ns-examples}, and \ref{bh-examples}, are possible. Nevertheless, it is important to note 
that ${\cal F}$ can also be much smaller, particularly for photometric
detections of lensing.     
   
\subsection{Identification of Potential Lenses} 

\subsubsection{White Dwarfs}

We identified nearby white dwarfs with measured distances and 
proper motions by using  the SUPERBLINK 
catalogue (L\'epine 2017, in preparation), which
employed the method outlined in \citet{1538-3881-145-5-136} to measure proper motions
larger than $40$~mas~yr$^{-1}$, and to estimate photometric parallax. 
The catalogue  identifies $35,246$ white dwarfs 
within $\approx100$ pc of the Sun. The highest rates 
associated with nearby 
white dwarfs, are illustrated in Table \ref{wd-examples}.  
 To compute $\theta_E$ we took the masses of the 
white dwarfs to be $0.6\, M_\odot.$ While the true mass distribution is
certainly more complex, actual values of $\theta_E$ will differ
from those we computed by a factor generally smaller than $\sim$ 0.4. 
Table \ref{wd-examples} shows that the white dwarfs in our sample are nearby.

\begin{table*}
	\centering
	\caption{Proper motion $\mu$, lens distance $D_L$, and Einstein
  angle $\theta_E$ of the most promising white dwarf lensing
  candidates from the SUPERBLINK proper motion catalogue.}
	\label{wd-examples}
	\begin{tabular}{lcccccc} 
		\hline
		Name & $\mu$ & $D_L$ & $\theta_E$ & Detectability & Density & Rate \\
		& arcsec yr$^{-1}$ & pc & mas & Factor & per arcmin$^{2}$ & per Decade\\
		\hline
		PM I11457-6450 & 2.69 & 3.9 & 33.15 & 110.50 & 38.83 & 2.12  \\
		PM I16233-5237 & 0.82  & 12.8 & 18.30 & 60.99 & 132.42 & 0.67 \\
		PM I12248-6158 & 0.91  &11.6  & 19.22 & 64.07 & 77.03 & 0.48 \\
		PM I18308-1301 & 0.91 & 11.6 & 19.22 & 64.07 & 73.85 & 0.46 \\
		PM I07456-3355 & 1.71 &  6.2 & 26.29 & 87.63 & 13.05 & 0.29 \\
		PM I13332-6751 & 0.80 & 13.3 & 17.95 & 59.84 & 53.16 & 0.25 \\
		\hline
	\end{tabular}
\end{table*}

\subsubsection{Neutron Stars}

The Australian Telescope National Facility (ATNF)\footnote{ATNF pulsar
  catalogue - \\ http://www.atnf.csiro.au/people/pulsar/psrcat/} ~pulsar
catalogue \citep{2005yCat.7245....0M} listed 2,536 pulsars as of
2016 January 31. From this list we selected all 242 pulsars with known
proper motions and estimated distances. We also added 6 well-known
neutron stars that also have proper motions and distances not
included in the ATNF, as well as two microquasars SS 433 and
Cygnus X-3 \citep{2014PASA...31...16M}. Overall this makes a list of
250 neutron stars with measured proper motions. 

For the purpose of estimating the value of $\theta_E$, we used
$M_{ns}\approx 1.4\, M_\odot$, which should generally produce
deviations from the true values of at most tens of per cent. Table
\ref{ns-examples} showcases candidate lenses from the neutron star
sample which combine large proper motions with
relative proximity to produce large event rates per lens.

\begin{table*}
	\centering
	\caption{Neutron stars that are good lensing candidates based
  on their proper motion $\mu$, lens distance $D_L$, and Einstein
  angle $\theta_E$.}
	\label{ns-examples}
	\begin{tabular}{lcccccc} 
		\hline
		Name & $\mu$ & $D_L$ & $\theta_E$ & Detectability & Density & Rate \\
		& mas yr$^{-1}$ & kpc & mas & Factor & per arcmin$^{2}$ & per Decade\\
		\hline
		J1741-2054    & 109.0  & 0.30 & 5.77 & 19.24 & 104.41 & 7.03$\times10^{-3}$ \\
		J1932+1059    & 103.4  & 0.31 & 5.68 & 18.93 & 78.62 & 4.86$\times10^{-3}$ \\
		J1856-3754    & 332.3  & 0.16 & 7.91 & 26.35 & 11.78 & 4.53$\times10^{-3}$ \\
		J0633+1746    & 178.2  & 0.25 & 6.32 & 21.08 & 9.87 & 1.30$\times10^{-3}$ \\
		J0835-4510    & 58.0  & 0.28 & 5.98 & 19.92 & 30.24 & 1.12$\times10^{-3}$ \\
		\hline
	\end{tabular}
\end{table*}

\subsubsection{Black Holes} 

We consider as potential lenses 60 black hole candidates
 from the BlackCat catalogue of
\citet{refId0}. Background stellar densities and
microlensing rates can be calculated for all of them. However, only five
of the black holes have measured proper motions, so detailed
trajectories and predictions can only be calculated for those five;
these are listed in Table~\ref{bh-examples}.
The range of possible black hole masses is large, extending
from a value that may be as low as $2\, M_\odot \-- 3\, M_\odot$ up to tens of solar masses.

\medskip 

\begin{figure}
  \centering \tiny
  \includegraphics[width=0.49\textwidth]{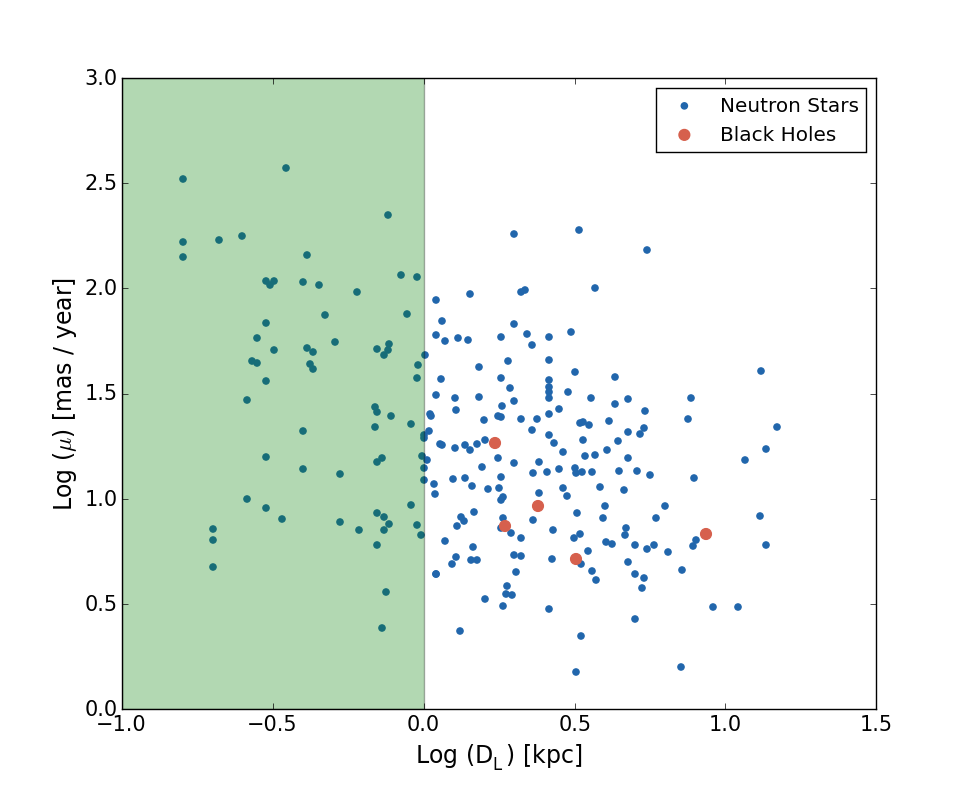}
  \caption[Steady flow]{Proper motions and distances
    for the neutron stars and black holes included in this study. 
    Their proximity and generally higher proper motion means that these
    objects can  generate lensing events at relatively high rates. 
	Proper motions
    are available at this time for only five black hole systems;
    they are shown in red.}
  \label{fig:ns-bh-mu-Dl}
\end{figure}

Fig. \ref{fig:ns-bh-mu-Dl} is a log-log plot of proper motion
versus distance for all the neutron stars and black holes we consider
as potential lenses. These two
are the most critical parameters for achieving a high microlensing
rate per lens.
Potential lenses located within a kpc will therefore dominate the
rate of events that can be predicted\footnote{Note however, that 
because there are many more
lenses in larger volumes, distant lenses produce more events overall. It is,
however, difficult to discover these more distant compact objects and it is
therefore not possible to make predictions of events to be generated by the 
large majority of them.}. 
The green shaded region highlights all objects located within 1~kpc. One
fourth of the potential lenses we consider
 are in this region, and 33 of them have
proper motions $\mu \geq 40$ mas yr$^{-1}$; 
These neutron stars and black holes produce the highest average microlensing
rate,
per lens.

\begin{table*}
	\centering
	\caption{Proper motion $\mu$, lens distance $D_L$, and Einstein
  angle $\theta_E$ for the five black holes in the BlackCat
  catalogue of \citet{refId0} that happen to have
  measured proper motions.}
	\label{bh-examples}
	\begin{tabular}{lcccccc} 
		\hline
		Name & $\mu$ & $D_L$ & $\theta_E$ & Detectability & Density & Rate \\
		& mas yr$^{-1}$ & kpc & mas & Factor & per arcmin$^{2}$ & per Decade\\
		\hline
		Cyg X-1       &  7.43  & 1.86 & 7.54 & 25.13 & 38.52 & 3.01$\times10^{-4}$ \\
		V404 Cyg      &  9.20  & 2.39 & 5.19 & 17.29 & 21.65 & 9.92$\times10^{-5}$ \\
		GRO J1655-40  &  5.19  & 3.20 & 3.66 & 12.20 & 58.25 & 7.49$\times10^{-5}$ \\
		GRS 1915+105  &  6.83  & 8.60 & 3.16 & 10.52 & 13.05 & 1.65$\times10^{-5}$\\
		XTE J1118+480 & 18.36  & 1.72 & 5.35 & 17.84 & 1.27 & 1.24$\times10^{-5}$ \\
		\hline
	\end{tabular}
\end{table*}

\subsection{Catalogues of Background Sources}

One reason that event prediction is now a practical endeavour is the
availability of a variety of all-sky catalogues which can be used to
map out background sources in the vicinity of known target lenses. For
the present study, we utilized the catalogues listed in Table
\ref{catalogues}, which list sources detected at optical wavelengths and
in the infrared. 

The \textit{Gaia} mission is of special importance to this study. First, 
\textit{Gaia} has the highest
astrometric precision of any all-sky survey to date, 
with the first data release
(September 2016) providing positions of 1 billion nearby stars to $10$ 
mas precision \citep{refId02}. The survey will have a magnitude 
limit of $20.5$ mag, and will eventually measure the positions of the 
brightest (dimmest) stars to about $5-10$ $\mu$as (a few hundred
$\mu$as). \textit{Gaia} will therefore discover many lensing events,
among them events in which some of the compact objects we consider here
serve as lenses. 

The Sloan Digital Sky Survey has lower astrometric precision  ($45$ mas RMS 
for each coordinate). It also has a fainter magnitude limit and can 
thus identify
dimmer background sources. Unfortunately, SDSS has limited
coverage at low galactic latitudes, where large surface densities
can assure higher rates of lensing events.  

With its sensitivity and high angular resolution, \textit{HST} can provide 
a unique look at the background of a potential lens. Source
extraction for many fields with
\textit{HST} images has been conducted and the results are listed in the 
The Hubble
Source Catalog (HSC), which we have used. The HSC
 reaches the faintest magnitudes with
relatively high astrometric precision, but it has very limited sky
coverage. 

The 2MASS catalogue is useful as an all-sky infrared catalogue,
particularly when the potential lenses are 
compact objects. White dwarfs,  
for example, tend to be relatively blue, 
and are therefore dim in the infrared. Blending
of light from the the lens is therefore less problematic
than it can be in other cases.

These 4 catalogues complement each other. In addition, they have some
stars in common, which allows  
cross-matching  
to establish a frame of reference that
improves the precision of relative positions. In addition, the fact
that the observations in the catalogues were generally taken at different
times,  
provides additional constraints
on the relative proper motions of the lens and background sources.

\begin{table*}
	\centering
	\caption{List of catalogues used to supply background
  sources in the fields of the lens candidates.}
	\label{catalogues}
	\begin{tabular}{lccr} 
		\hline
		Catalogue & Waveband & Magnitude Limit & Reference\\
		\hline
		2MASS & Infrared &  J$\approx$16 & \citet{2003tmc..book.....C}\\
	    Hubble Source Catalog (HSC) & Optical/Infrared  & V$\approx$24 & \citet{2015AAS...22533205W}\\
		Gaia first data release & Optical & V$\approx$20 &\citet{refId3}\\
		SDSS & Optical & V$\approx$22 & \citet{2012ApJS..203...21A}\\
		\hline
	\end{tabular}
\end{table*}

\subsection{Background Source Densities}

The 2MASS, SDSS, and Gaia catalogues are all available for query through
the VizieR
service\footnote{VizieR - http://vizier.cfa.harvard.edu/viz-bin/VizieR}. 
For each catalogue,
we retrieved all sources located within 1 arcmin of the
position of each of the the $2,536$ neutron stars listed in the ATNF, each
of the $60$
black holes from the BlackCat catalogue, and each of the $35,246$ white dwarf
candidates from the SUPERBLINK catalogue. We then calculated background
source densities per square arcmin in the 
vicinity of each compact object by
counting the number of sources within 1 arcmin
of it, then dividing by the area ($\pi$ square arcminutes). 

\begin{figure}
  \centering
  \includegraphics[width=0.45\textwidth, height =55mm]{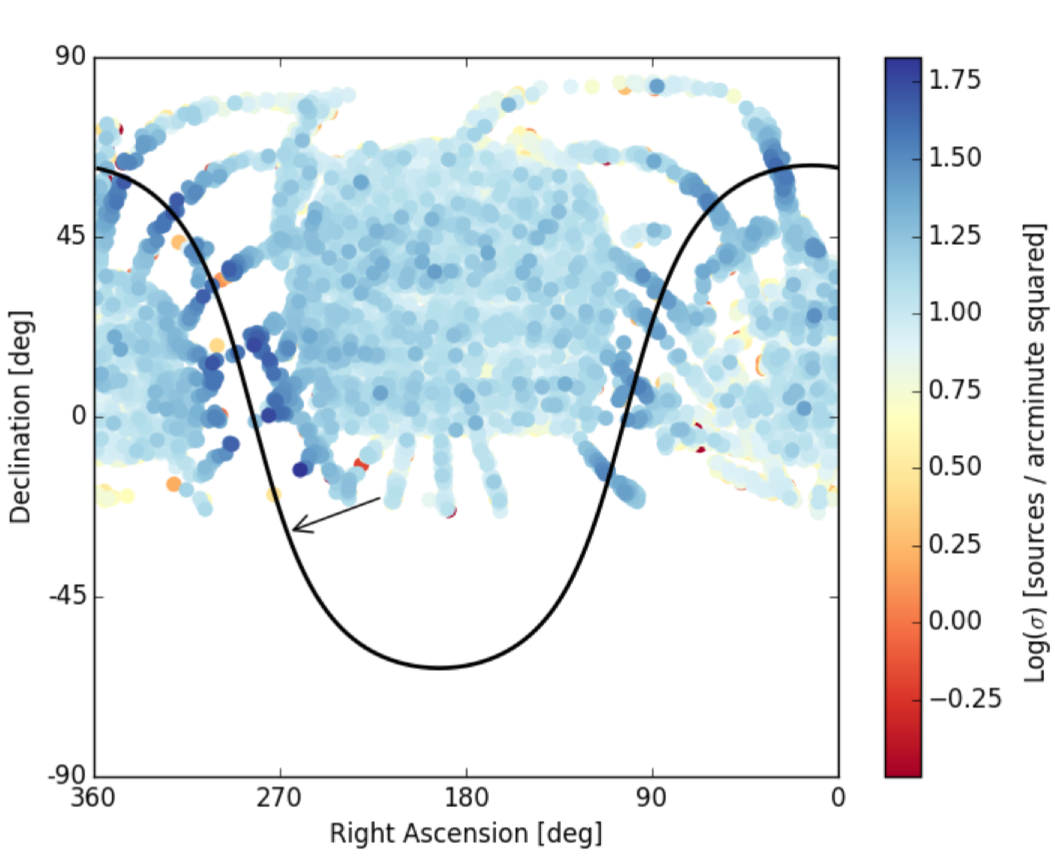}
  \includegraphics[width=0.45\textwidth, height =55mm]{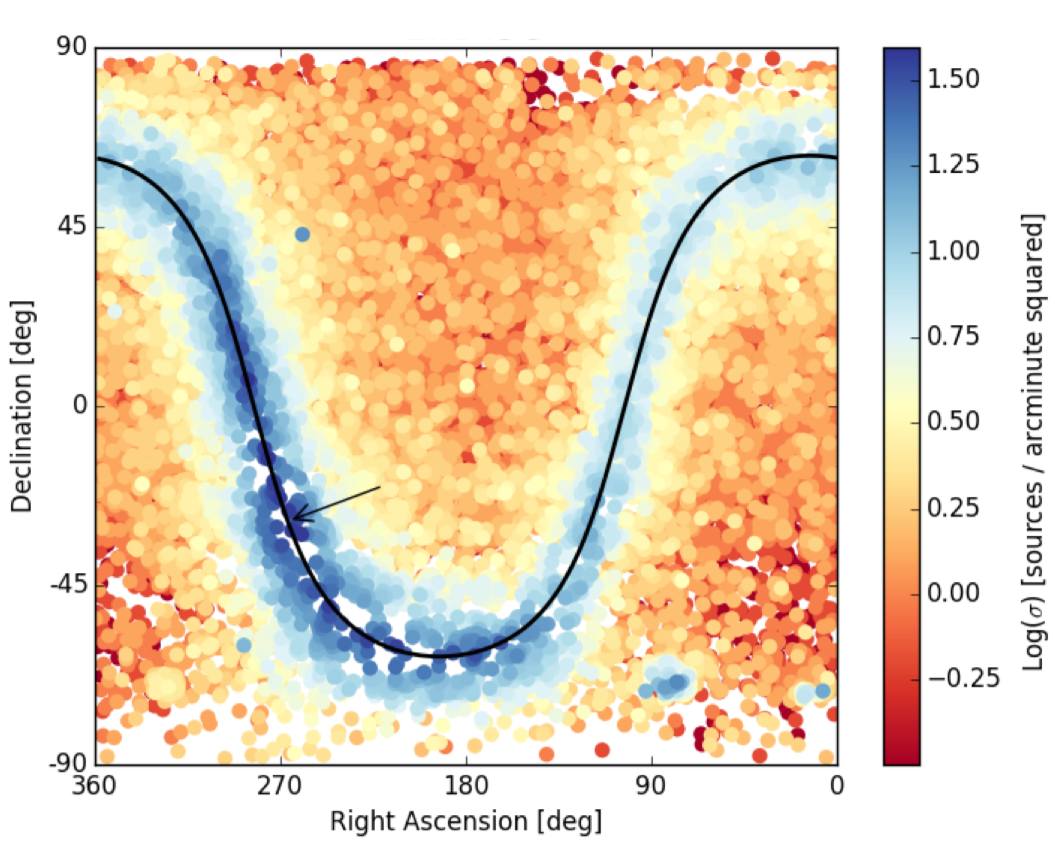}
  \caption[Steady flow]{Density of background sources in the areas
    immediately surrounding the 35,246 white dwarfs considered in this
    study. Each coloured orb represents one of the white dwarfs, with
    the colour corresponding to the source field density. These two
    panels show the field density for sources listed in 
    SDSS and 2MASS. 
The colour scales with the log of the
    local field density; note that the colour scale varies between
    plots. The densest regions appear along the black line that
    represents the galactic plane. The arrow in each map points to the
    location of the galactic center.}
  \label{fig:wd-densities}
\end{figure}

\begin{figure}
  \centering
  \includegraphics[width=0.45\textwidth, height =55mm]{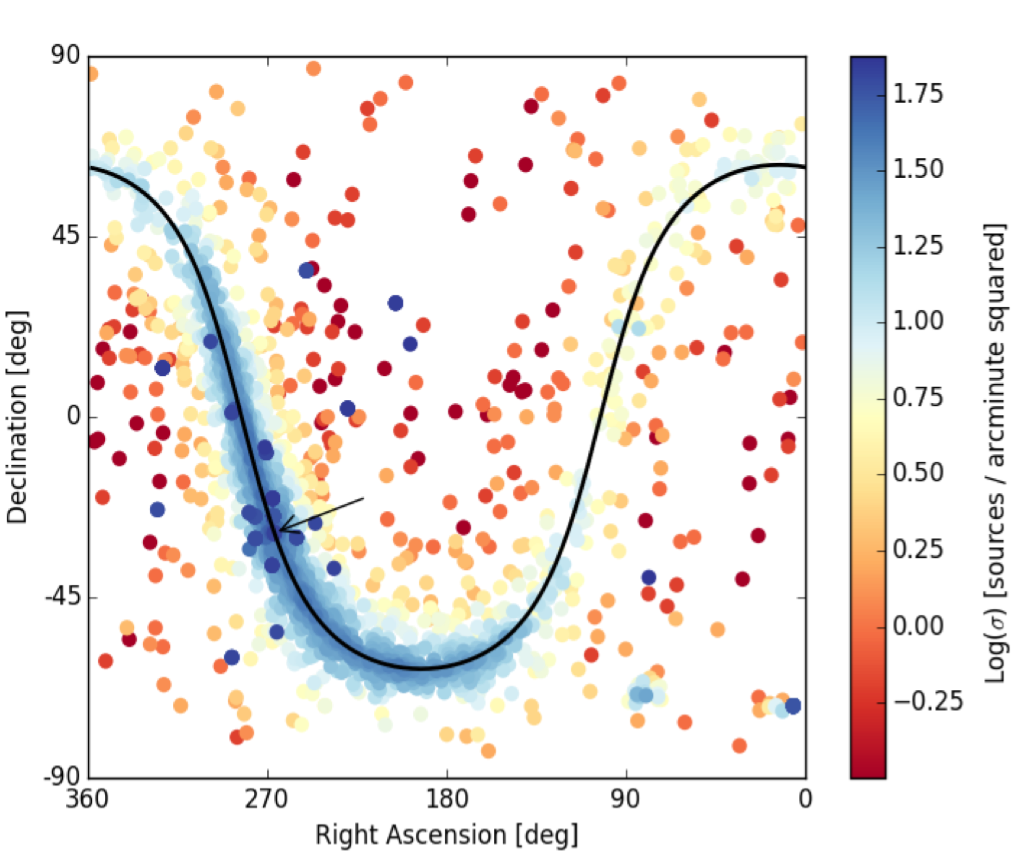}
  \caption[Steady flow]{Density of background sources in the areas
    immediately surrounding the $2,544$ neutron stars considered in this
    study. Each coloured orb represents one of the neutron stars,
    with the colour corresponding to the source field density. The 
    panel shown here maps the field density for sources listed in 
    2MASS. 
 We expect neutron stars near the
    bulge to have the highest probability of generating microlensing
    events due to the large number of sources in this region.}
  \label{fig:ns-densities}
\end{figure}

Fig. \ref{fig:wd-densities} shows the SDSS and 2MASS
density maps for the white dwarfs
and Fig. \ref{fig:ns-densities} shows the 2MASS
background source density map
for the neutron stars. As expected, the densest regions
follow the line of the galactic plane, with some of the highest densities found 
near to the galactic bulge. Regions of high source
density outside the Disk include the Magellanic Clouds and some
globular clusters. 

We computed densities in background fields for 
the 26
neutron stars and 1 black hole with recorded proper motion that had been imaged
by \textit{HST} before early 2016.
We also required that there be a corresponding source list
in the HSC. We used the interactive viewer in the Hubble Legacy
Archive\footnote{Hubble Legacy Archive - hla.stsci.edu}. 
We calculated the
field densities by counting all HSC-listed sources in those \textit{HST} images,
dividing by the coverage area of the particular imager that was used
(WFPC2, ACS, or WFC3). In cases of observations with multiple
overlapping fields, we counted only the sources within the borders of
the field whose center was closest to the target. If the sources were
extracted from a WFPC2 image, we also excluded sources from the PC
chip. The resulting densities are shown in 
Table \ref{ns-hsc-densities}. As expected,
source densities are systematically higher than those from
the ground-based catalogues because of the fainter magnitude limit. 

\begin{table*}
	\centering
	\caption{Background source densities in the Hubble Source
  Catalog, for neutron stars that have been imaged with \textit{HST}. The table
 also lists the ratio between the HSC source density and the values calculated from the other three catalogues (2MASS, SDSS, Gaia) to show  the significant increase in background sources that can be achieved
with deeper surveys.}
	\label{ns-hsc-densities}
	\begin{tabular}{lcccc}
		\hline
		Lens ID & HSC source density &$\epsilon_{SDSS}$ & $\epsilon_{2MASS}$& $\epsilon_{Gaia}$\\
		& arcmin$^{-2}$  &  &  &\\
		\hline
		J0205+6449 & 110.01& 2.93& 16.46& 5.57\\
J0437-4715 &   6.61&    -& 10.39& 3.46\\
J534+2200 & 149.76 &  -  & 10.69 & 20.46 \\
J0621+1002 & 96.0 &- & 15.87 & 8.38 \\
J0633+1746 & 35.63 &- & 6.22& 3.61 \\
J0659+1414 & 50.56 & 2.48 & 13.24& 8.36 \\
J0821-4300 & 106.45&    -& 19.67 & 6.43\\
J0823+0159 & 10.67& 0.88&  8.38&  4.79\\
J0835-4510 & 205.65 & -& 20.84 & 6.80\\
J1057-5226 & 128.0 & - & 14.36 & 6.09\\
J1455-3330 & 89.39&    -& 70.20 & 17.55\\
J1640+2224 & 25.17& 2.26& 39.54 & 9.89\\
J1713+0747 &  84.91&    -& 88.91& 26.67\\
J1856-3754 & 199.04&    -& 26.05& 16.90\\
J1857+0943 & 391.47&    -& 18.92& 6.58\\
J1952+3252 & 583.04&    -& 26.17& 9.49\\
J1959+2048 & 299.09&    -& 19.18& 5.22\\
J2019+2425 & 138.24&    -& 18.10& 4.34\\
J2033+1734 &  93.23&    -& 26.63& 8.37\\
J2051-0827 &  23.68&    -& 14.88& 9.30\\
J2145-0750 &  46.51& 6.64& 73.05& 29.22\\
J2225+6535 &  54.40&    -& 11.39& 5.70\\
		\hline
		Average $\epsilon$ &  & 3.04 & 25.87 & 10.14\\
		\hline
	\end{tabular}
\end{table*}

\begin{figure*}
  \centering
  \includegraphics[width=0.8\textwidth, height= 70mm]{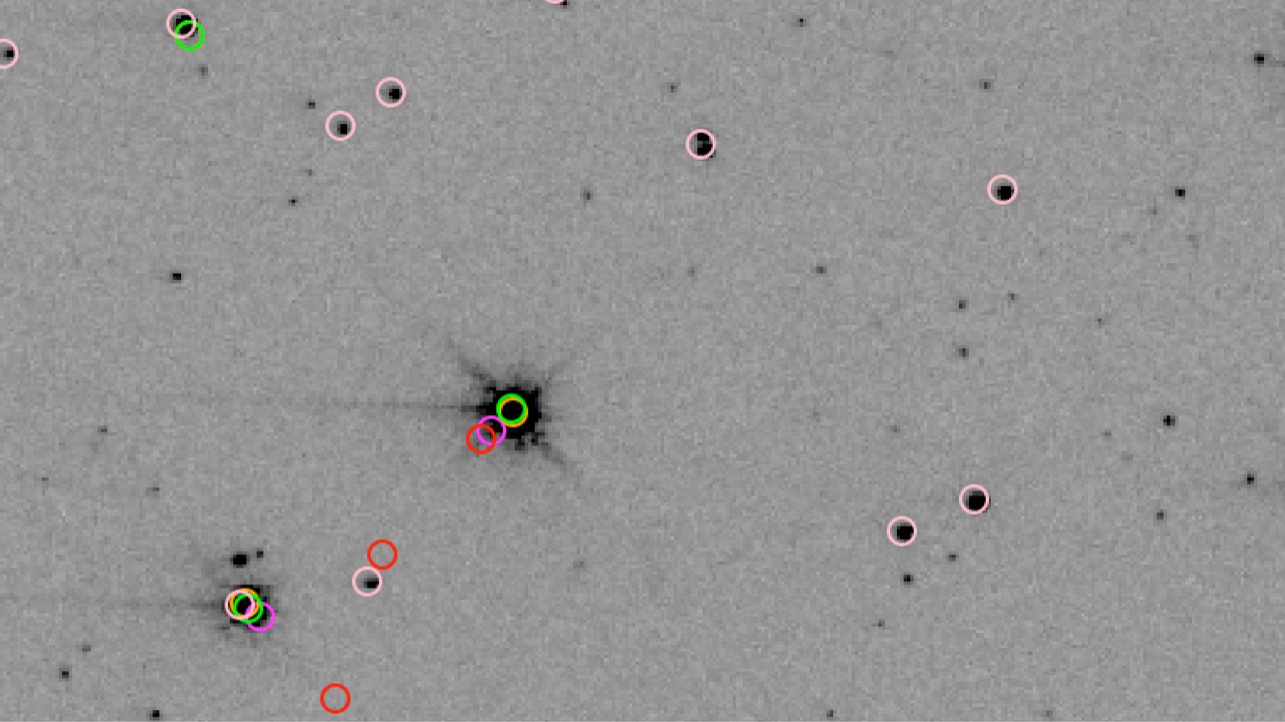}
  \caption[Steady flow]{\textit{HST} WFPC2 image taken in the F675W filter on
    the 20$^{th}$ June 2001. The bright companion of the black hole
    GRO J1655-40 can be seen near the center of the field. The sources
    circled are those found in the source catalogues we've used: HSC
    (pink), GSC2.3 (purple), Gaia (orange), 2MASS (green) and
    USNO-B1.0 (red). Many sources are visible in the image that
    have not been included in any of the source catalogues.}
  \label{GRO}
\end{figure*}

We note that source lists in the HSC are not necessarily complete. 
One example is shown in
Fig. \ref{GRO} which plots an observation of GRO J1655-40 with the WFPC2 on \textit{HST}. The overlay of sources listed in the
HSC clearly only flag the brightest objects in the field, missing the
majority of the fainter sources, which tend to be $\sim10$ times more
numerous than those listed in the HSC. Because most of the unlisted
stars are dimmer, approaches by a passing stellar remnant
may have to be closer in order for an event to be detectable.
Nevertheless, this suggests that 
the rate of microlensing events detectable through
\textit{HST} imaging may be several times  
higher that the rates predicted based on the HSC alone.
It is also important that even background sources that are distant galaxies
can be detectably lensed, but these are not included in the HSC.

Only a small fraction of the sky has been imaged by \textit{HST}. Yet, since 
it can be a frontline tool for the prediction and observation 
of astrometric lensing events,
it is useful to
estimate the background densities that could be measured, were \textit{HST} 
employed to image more fields containing stellar remnants. 
We have therefore considered fields for which we have computed 
background surface densities based on SDSS, 2MASS, and/or Gaia,
and for which \textit{HST} imaging is also available. In the latter case, we
have used the HSC to compute the local source density.
The specific fields we have studied in this way all contain the coordinates of
a neutron star with known distance and proper motion. For each
such field, we have $\sigma_{\rm HSC}$ from the HSC, and $\sigma_{\rm cat}$
from some other catalogue(s). We define the ratio  
$\epsilon = \frac{\sigma_{\rm
    HSC}}{\sigma_{\rm cat}}$. Average values $\epsilon$ for all
fields are found to vary between $3-25$. This means 
that, by taking \textit{HST} images of regions with potential lenses in the foreground,
we can increase the 
rate at which we can observe lensing events 
by around an order of magnitude.

\subsection{Microlensing Rates}

We apply Equation \ref{eqn:rates} to calculate the average
rate of microlensing events for each candidate lens, using
the measured background source densities estimated above. We start by considering individual compact objects which, because some combination of proximity and high proper motion, may produce microlensing events at high average rates. Examples are shown in Tables \ref{wd-examples}, \ref{ns-examples}, and \ref{bh-examples}, for white dwarfs, neutron stars, and black holes respectively. Note that the rates are averages, based on the local background density. 

To determine the time of the next event for any individual compact object, it is necessary to consider the positions of each nearby background source with respect to the path of the compact object. Nevertheless, this short list
makes it clear that it is not difficult to identify individual white dwarfs that will produce
events in the near future. For neutron stars and black holes, the
rates are far smaller. Nevertheless, in \S 5 we will show through
the study of a specific neutron star, that predictions may be
possible for small numbers of neutron stars, and potentially
for black holes as well.

We first calculate individual rates for the $35,246$ white dwarfs for each of the 
photometric catalogues, shown in Table \ref{wdsb-densities}. The top of Table \ref{wdsb-densities} lists the rate per white dwarf per decade.   
The fourth column suggests that one would need to study the
positions of approximately $1000$
white dwarfs in order to detect an event each decade. We note however, that
these estimates are based on averaging white
dwarfs that produce events several times per century with others  
that produce events at much lower rates. It is therefore possible to
 select a smaller subset of white dwarfs to ensure detecting one or
more events per decade.

It is also important that large values of the detectability
factor, ${\cal F}$, mean that the time interval during
 which astrometric shifts can be
detected may be years or decades. If, therefore, we identify
our computed rates with the numbers of events that will
peak per year, the numbers that will be detectable each year    
are larger, because some detectable events will have peaked 
during previous decades
and others will peak during future decades. 

\begin{figure}
\centering
\includegraphics[width=0.49\textwidth]{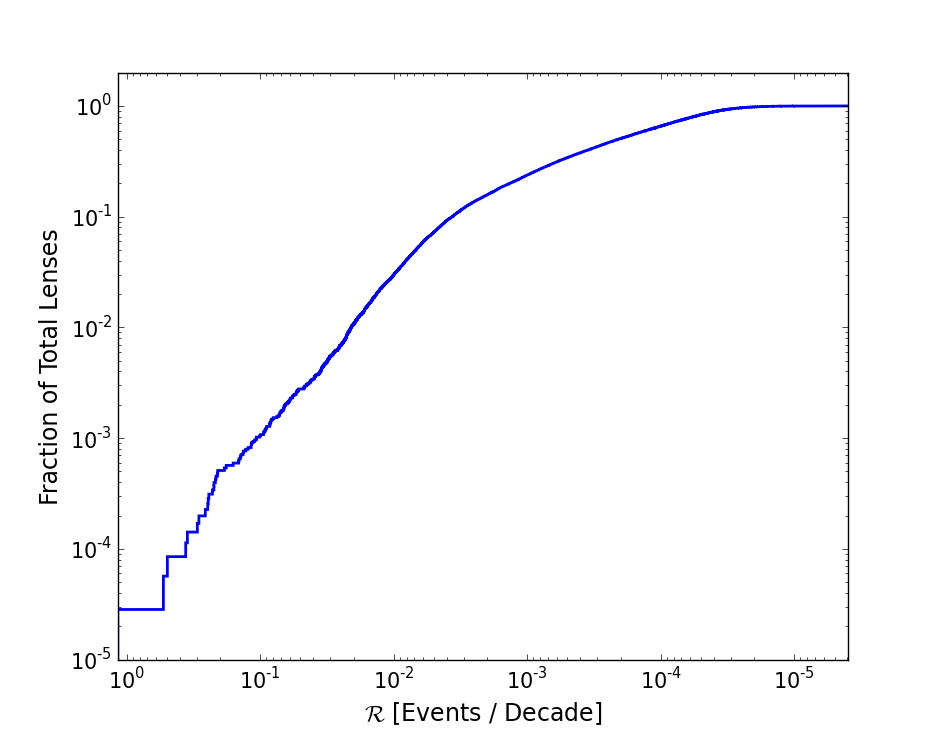}
\includegraphics[width=0.49\textwidth]{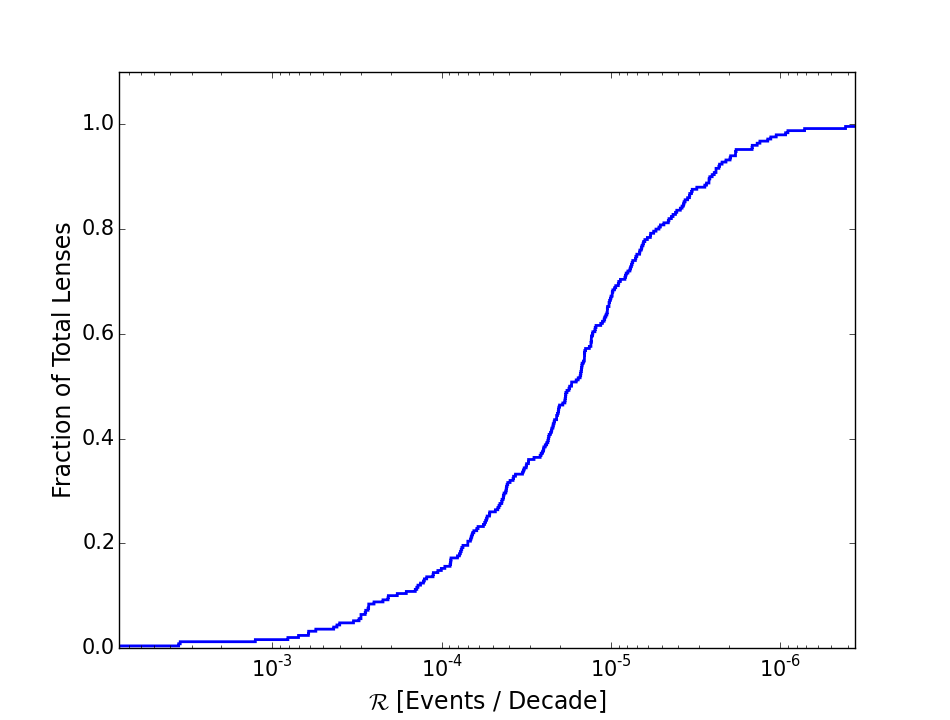}
\caption{Cumulative distribution plots for the the event rate, $\cal{R}$, for the white dwarfs ({\sl Top}) and neutron stars ({\sl Bottom}) with measured
  distances and proper motions. Plotted on the horizontal
axes is the logarithm to the base ten of the rate per decade calcualted using $\sigma$ from the \textit{Gaia} catalogue.
  }
\label{wd_selection}
\end{figure}

Fig. \ref{wd_selection} shows two cumulative distribution plots, with the estimated event rate, $\cal{R}$, in a logarithmic scale on the x-axis. The rate values are those determined with background source densities measured from the \textit{Gaia} catalogue. The top panel in the figure is for the white dwarfs which shows that $\sim$ $0.1$ per cent of the white dwarfs ($\sim$ $35$ lenses) are expected to produce events at a rate $> 0.1$ events per decade. These white dwarfs are the most interesting to focus on because they have a good combination of high proper motions, high density background stellar fields, and are at relatively small distances from Earth.

In the bottom half of Table \ref{wdsb-densities} the total combined rates for all
of the white dwarfs we consider are listed. Even without the \textit{HST} enhancement factor,
and even before considering events that may peak during prior or
subsequent decades, 
we expect $2-4$ dozen white-dwarf-generated astrometric events per decade. Note that these total combined rates refer only to the total set of white dwarfs that we have considered. By including known white dwarfs with smaller measured proper motions, the numbers would be significantly increased. Futhermore, ongoing studies, including those with {\it Gaia}, are adding to the list of white dwarfs with measured distances and proper motions.

We have carried out the same procedure for the $250$ neutron stars and $5$ black
holes in our sample. We averaged the
values to estimate an overall rate of microlensing events per
lens. These rates are calculated separately for each photometric
catalogue, and listed in Tables \ref{ns-density} and \ref{bh-density}. The rates quoted in the top portions of the tables are expressed in microlensing events per compact object per decade. 
The numbers are roughly ten times smaller for neutron stars (relative to 
white dwarfs), 
because there are so few neutron stars with measured distances and proper
motion, that the 3-dimensional density is small and most
are at larger distances and have smaller proper motions than a
large fraction of the white dwarfs in our sample. 
There is an additional factor-of-ten
decline for black holes. 

The bottom panel of Fig. \ref{wd_selection} shows the cumulative distrubtion plot for the event rates of the neutron stars. It can be seen in the figure that the top $\sim$  $15$ per cent of the neutron stars are estimated to only produce $> 10^{-4}$ events per decade. The message from these calculations is
that we would need to identify more neutron stars and black holes 
and measure their distances and proper motions, in order
to make event prediction a practical endeavour. In spite of this apparently
grim prognosis, we will show in \S 5 that at least one neutron star
has apparently generated events in the recent past.  
This demonstrates that fluctuations can
play a role in increasing the local rate of events (as opposed to
the long-term average). 
In addition, 
the existence of dim background stars that have 
not been catalogued in the HSC mean that our computed rates are too low. The
example we present in \S 5 explicitly illustrates this point.

Despite the fainter magnitude limit of the SDSS, the event rate from
this catalogue is comparable to the event rate for the 2MASS catalogue,
and about an order of magnitude lower than for the Gaia catalogue. This
is because of the patchy coverage of SDSS at low magnitude, which
excludes the densest areas of the sky. This shows the need for deeper
astrometric surveys at low galactic latitudes, which would
significantly increase the potential for event prediction.

\begin{table*}
	\centering
	\caption{Estimated average event rates for the white dwarf
  candidates, calculated from the surface densities
 of the background regions of white dwarfs with measured
  proper motions. Top: the rate per compact object, averaging over all the
white dwarfs in our sample.
Bottom: the overall rate is the rate
  expected from all the white dwarfs in our sample combined.}
	\label{wdsb-densities}
	\begin{tabular}{lcccc} 
		\hline
		 Lens Category & Source  & Lenses with Nearby & Rate per Compact & Enhanced Rate\\
		& Catalogue & Background Sources & Object per Decade & (Rate x $\epsilon$) \\
		\hline
		White Dwarfs &SDSS & 18,208 & 1.61 $\times10^{-3}$ & 4.89 $\times10^{-3}$ \\
(35,246 objects)&2MASS &31,769 & 7.62 $\times10^{-4}$ & 1.97  $\times10^{-2}$ \\
&Gaia & 34,519    &  1.52 $\times10^{-3}$&  1.54 $\times10^{-2}$\\
		\hline
		\hline
		Lens Category & Source  & Lenses with Nearby & Overall Rate & Enhanced Rate\\
		& Catalogue & Background Sources & per Decade & (Rate x $\epsilon$) \\
		\hline
		White Dwarfs &SDSS & 18,208 & 29.27 & 88.98 \\
(35,246 objects)&2MASS &31,769 & 24.22 & 626.57  \\
&Gaia & 34,519   & 52.37 & 531.03 \\
		\hline
	\end{tabular}
\end{table*}

\begin{table*}
	\centering
	\caption{Estimated theoretical event rates for the neutron star
  and lens candidates, calculated from areas that contain 
  lenses with measured proper motions. The overall rate is the rate
  expected from all the lenses combined.}
	\label{ns-density}
	\begin{tabular}{lcccc} 
		\hline
		Lens Category & Source  & Lenses with Nearby & Rate per Compact & Enhanced Rate\\
		& Catalogue & Background Sources & Object per Decade & (Rate x $\epsilon$) \\
		\hline
		Neutron Stars  & SDSS  &  54 & 1.01$\times10^{-4}$ & 3.07$\times10^{-4}$ \\
(250 Objects)  & 2MASS  & 239 & 5.31$\times10^{-5}$ & 1.37 $\times10^{-3}$ \\
 & Gaia       & 249 & 1.36$\times10^{-4}$ &  1.38$\times10^{-3}$\\
		\hline
		\hline
		Lens Category & Source  & Lenses with Nearby & Overall Rate & Enhanced Rate\\
		& Catalogue & Background Sources & per Decade & (Rate x $\epsilon$) \\
		\hline
		Neutron Stars  & SDSS  &  54 & 5.43$\times10^{-3}$ & 1.65$\times10^{-2}$ \\
(250 Objects)  & 2MASS  & 239 & 1.27$\times10^{-2}$ & 3.29$\times10^{-1}$ \\
 & Gaia       & 249 & 3.38$\times10^{-2}$ &  3.43$\times10^{-1}$\\
		\hline
	\end{tabular}
\end{table*}

\begin{table*}
	\centering
	\caption{Estimated theoretical event rates for the black hole lens candidates, calculated from areas that contain lenses with measured proper motions. The overall rate is the rate
  expected from all the lenses combined.}
	\label{bh-density}
	\begin{tabular}{lcccc} 
		\hline
		 Lens Category & Source  & Lenses with Nearby & Rate per Compact & Enhanced Rate\\
		& Catalogue & Background Sources & Object per Decade & (Rate x $\epsilon$) \\
		\hline
		Black Holes & SDSS      & 1 & 4.03$\times10^{-5}$ & 1.23$\times10^{-4}$ \\
(5 Objects)   & 2MASS     & 5 & 5.36$\times10^{-5}$ & 1.39 $\times10^{-3}$ \\
 & Gaia     & 5 & 1.01$\times10^{-4}$ & 1.02$\times10^{-3}$\\
		\hline
		\hline
		Lens Category & Source  & Lenses with Nearby & Overall Rate & Enhanced Rate\\
		& Catalogue & Background Sources & per Decade & (Rate x $\epsilon$) \\
		\hline
		Black Holes & SDSS      & 1 & 4.03$\times10^{-5}$ & 1.23$\times10^{-4}$ \\
(5 Objects)   & 2MASS     & 5 & 2.68$\times10^{-4}$ & 6.93 $\times10^{-3}$ \\
 & Gaia     & 5 & 5.04$\times10^{-4}$ & 5.11$\times10^{-3}$\\
		\hline
	\end{tabular}
\end{table*}


\section{Predicting Close Passages}

In \S 3 we computed the  average rate of events
to be generated by each lens  using the density, $\sigma$, of background
stars. 
Typical densities are small enough that local 
fluctuations actually determine when the next detectable
 event will be generated
by
a specific foreground lens. For example, a potential lens producing
a low average rate may happen to be close to a nearby star,
and it may also be traveling toward the position of that star.
Such a mass may therefore produce a detectable lensing event in the near future.
In this section we therefore consider the individual 
position of each background star
listed in each catalogue. 
Using the position of the foreground object as it was during
a specific epoch, and its proper
motion, we determine the distances and times of specific 
future close approaches
it will make to background stars.
This procedure identifies the potential lenses whose paths
should be studied in the near future. 
It is the next step toward our
ultimate goal of 
predicting 
specific lensing events to be
generated by compact objects, and planning observations of the associated
centroid shifts. 

\subsection{General Method and Caveats}

We consider the sets of white dwarfs, neutron stars and black holes
introduced in \S 3. For each nearby stellar remnant with known proper
motion, we retrieve the positions of background field stars.
Because future observations of astrometric events will likely be taken
by \textit{HST} or \textit{Gaia}, we explore the close approaches between
potential lenses and the stars listed in the HSC and in the Gaia catalogue. We
focus on approaches that would cause centroid shifts
$\delta(u)>0.3$ mas, a realistic limit for \textit{HST}.
We note, however, that \textit{Gaia} can measure positions with higher
precision, making it possible to detect events for more distant approaches
and thereby increasing the event rate. 

We then calculate the angular distances
between the background sources and a straight line representing the
extrapolated path of the stellar remnant over 
time.
We flag any
approaches that come close enough for a detectable lensing event to
occur, and calculate both the time of closest
approach and the angular separation at that time.
If this procedure identifies one or more close approaches,
parallax effects and the motion of the background star must also be
computed and taken into account.  

In practice, the calculations are affected by several sources of
uncertainty: (1) uncertainties in the absolute positions of faint background
sources; (2) uncertainties in the proper motions of the potential lenses, which can build to uncertainties of $\sim1$ arcsec over a century; (3) the proper motions of typical field stars, which may be a few mas yr$^{-1}$ due to random motions or to galactic differential
rotation; these can add uncertainties of $\sim1$ arcsec over a
century. These three factors contribute to the uncertainty in such a way that
it accumulates over time, making reliable prediction of future events
strongly dependent on the predicted time of closest approach: the
sooner the event occurs, the greater the reliability.

On the other hand, 
even more distant approaches may have the potential to
produce microlensing events. For example, 
astrometric effects can be detected out to
$\theta_E^2/\delta\phi$, which can be larger than an arcsec. In addition,
the planetary system of a compact object may include wide-orbit planets,
possibly even to the  
edge of the Oort Cloud. Oort Cloud radii are in the range
$10^4\--10^5$~AU, and the distance to nearby stellar remnants are in the
range from about 10~pc to a few kpc. Thus, wide-orbit planets could
be $1\--1000$ arcsec from the stellar remnant they orbit.
 Note that, if we consider lensing by objects in the most
distant reaches of possible Oort Clouds,
then the positions of large numbers of background 
stars would need to be monitored. Here we will specifically consider only  
close approaches between the compact object itself and a background star.

\subsection{Computing the Path of the Lens}

For every object with a measured proper motion, $\mu$, we use the
components of the proper motion in right ascension and declination,
$\mu_{\alpha}$ and $\mu_{\delta}$ multiplied by $(t-t_{epoch})$.

\begin{equation}
\alpha (t) = \alpha_{epoch} + \mu_{\alpha}(t-t_{epoch}),  \quad
\delta (t) = \delta_{epoch} + \mu_{\delta}(t-t_{epoch}) 
\label{increment}
\end{equation}

We have computed the paths of all white dwarfs from the year 
2020 through the year 2035,
to allow us to identify any events of the recent past (which may be detectable, e.g. in
archived \textit{HST} data) as well as within the coming 
$\sim18$~years\footnote{In a separate paper we will
make specific  predictions for white-dwarf generated events,
and will consider times earlier than 2020. In the present paper we choose to
avoid the problem of self-identifications.
}
Because the numbers
of white dwarf we are considering is large, 
we will derive an 
estimate of the average event rate, which can be checked against the
rates computed in \S 3. 
Because there are many fewer neutron stars and black holes, 
we have projected their paths over a baseline that is $\sim200$ times longer, 
extending 2000 years
into both the past and future. 
We note that some events that occurred over historical times may 
have produced high enough magnifications to have been detected as naked
eye brightening of background stars. For such possible close
approaches, historical records, including
optical plates from the 1800s, can be investigated for
strong lensing events.

For each compact object, we retrieve from the relevant catalogue the
coordinates and magnitudes of all the sources located within 10 arcsec
of the line segment defining the path of the lens.

\subsection{Measured Approaches and Event Rates for White Dwarfs}

Our simulation of the paths of the white dwarfs from SUPERBLINK
 during the
interval 2016$\--$2035 found that every white dwarf will travel through a
region containing background stars listed in the Gaia catalogue, and
that $358$ approaches will produce detectable
astrometric lensing events.
This corresponds to an event rate of approximately 36 per decade,
somewhat lower than the average listed in Table~6. 
Roughly 46 per cent of these approaches will be closer than $10\, \theta_E,$
and 13 \wdf s will come within $\theta_E$ of a Gaia-listed star. 
High-resolution images of
these $13$ regions can determine whether the passages will actually
be this close. If so, detectable photometric effects are expected.
Only $62$ ($0.18$ per cent) of the white dwarfs will follow paths in regions
containing sources listed in the HSC. 
We found $31$ close approaches between \wdf s and HSC sources.

\subsection{Measured Approaches and Event Rates for Neutron Stars and
  Black Holes}

The HSC includes observations covering the fields of 26 of the neutron
stars. Over the time span under consideration ($\pm2000$ years), we
identify 289 approaches where any one of these 26 neutron stars with
come within $1$ arcsec of a background source. Of these, we find $15$
events with causing centroid shifts 
$\delta(u)>0.3$ mas. This suggests a rate of 
$\approx1.5 \times 10^{-3}$ per star per decade.

The Gaia catalogue lists sources along the paths of all $249$ neutron stars in
our sample. Over the time span under consideration ($\pm2000$ years),
we identify 338 approaches where any one of these 250 neutron stars
come within 1arcsec of a background source. Of these, we find $19$
events with centroid shifts 
$\delta(u)>0.3$ mas. This suggests a rate of
$\approx1.9 \times 10^{-4}$ per star per decade. The lower rate of occurrence
reflects the brighter magnitude limit of the Gaia catalogue (V$\lesssim$20) compared to that of the HSC (V$\lesssim$25).

\begin{figure}
  \centering
  \includegraphics[width=0.49\textwidth]{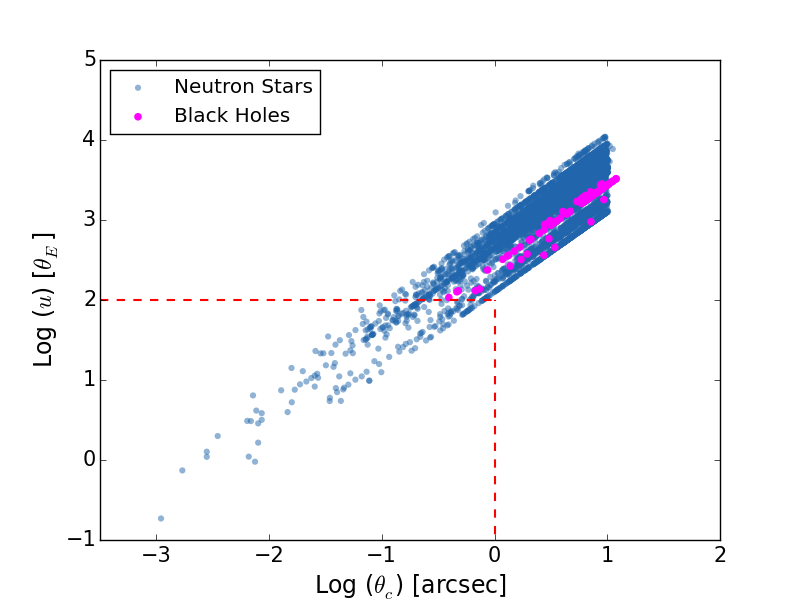}
  \includegraphics[width=0.49\textwidth]{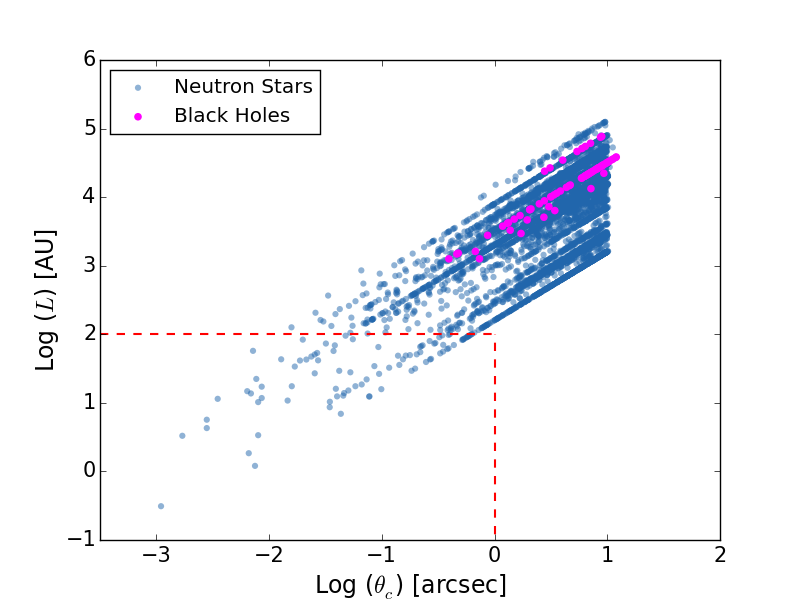}
  \caption[Steady flow]{{\sl Top:} log of the distance of closest
    approach in units of the Einstein angle plotted against the log of
    the impact parameter in arcsec, for all the potential
    lens/source encounters in our sample of neutron stars and black
    holes. Each oblique line artefact corresponds to a specific
    neutron star or black hole lens, because the Einstein angle and
    impact parameter are both correlated, and their ratio corresponds
    to our distance to the lens. Events for which astrometric shifts
    can be detected are those with distances of closest approach less
    than $100~\theta_E$ (horizontal dashed line) which happen only in
    events where the impact parameter is less than 1arcsec
    (vertical dashed line).
    {\sl Bottom:} physical distance of closest approach, in AU,
    plotted against the log of the impact parameter in arcsec, for
    the same subset.}
  \label{ns_plots}
\end{figure}

All the approaches from the analysis for the black holes and neutron
stars are plotted in Fig. \ref{ns_plots}. The top and bottom
plots show the approach distances in arcsec against the impact
parameter and distance $L$ (units of AU) respectively. The dashed red
lines border the zones of the most interesting approaches. Approaches
that come within $u=100~ \theta_E$ are potentially capable of
producing measurable lensing effects. The approaches with minimum
distances less than  $L=100$ AU provide good opportunities to
investigate the existence of close orbiting exoplanets.

We find that none of the approaches between the black holes and
background sources produced a maximum centroid shift $>0.3$ mas. Over
the $4,000$ year time period, $9$ approaches came within $1$ arcsec, none
of which are estimated to be in the detectable range. These rates were
to be expected, as $\ll 1$ event per decade was predicted in \S 3, even when multiplying by the HSC density factor $\epsilon$. With so
few objects in this sample, finding frequent events was unlikely. All
of these objects are several kpc away and have low proper motions
($\sim 10$ mas yr$^{-1}$) so they will only cover a small area of the sky,
even over long time periods.

\begin{figure}
  \centering
  \includegraphics[width=0.49\textwidth]{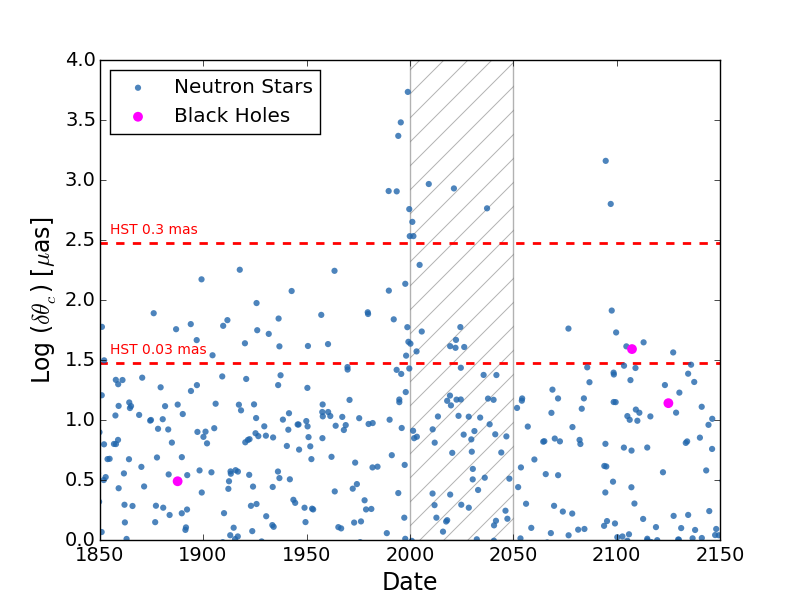}
  \caption[Steady flow]{A plot showing the log of the centroid shift
    at the closest approach for every passage made by neutron stars
    and black holes to nearby background sources. The two red lines
    show the detectability limits discussed in \S 1. The hashed
    region delineates the 2000$\--$2050 time period.}
  \label{detectables}
\end{figure}
The close approaches predicted by our simulation are based on a
detectability cut-off of $\delta(u) > 0.3$ mas. However if future
imaging techniques are improved (e.g. \citet{Riess:2014uga}) then more
approaches will fall into the detectable category. This is shown in
Fig. \ref{detectables} which plots the log of the maximum centroid
shift against the time of closest approach. The two red lines
correspond to detectability limits of $0.3$ mas and $0.03$ mas. From
this plot we can see that many more approaches would be detectable if
astrometric precisions of $x=0.03$ mas are attainable. 

\section{Prediction of Specific Events}

To predict a specific event, we need to estimate
the distance of closest approach in units of $\theta_E.$
Using images of the region, we can determine the distance of
closest approach in milliarcsec. To express the distance in units of
$\theta_E$, we need the distance, $D_L$ to the lens and
an estimate of its mass.
As before we use 
$M_L=1.4\, M_\odot$ for neutron stars and $M_L=0.6\, M_\odot$ for white dwarfs. 
If the lens produces an event that we can detect and study, we will measure its
mass to high precision.  

In principle, event prediction requires at least two high-resolution
images, to allow us to locate the potential lens relative to the
positions of background stars, and to measure relative motions.
We note, however, that in some cases we may find in a single high-resolution
image that the lens is already close enough to a background star to be
causing a detectable astrometric shift. When this occurs, follow-up
images (which may be taken several times over an interval of months or years)
can track the shift; a subsequent image can determine the unshifted
source position.  

\subsection{RX J1856.5-3754}

The \ns\ predicted to produce lensing events at one of the highest rate is 
RX J1856.5-3754. Known as a member of the Magnificent Seven
\citep{2007Ap&SS.308..181H}, RX J1856.5-3754 was one of the first
isolated neutron stars discovered by the ROSAT telescope. 
RX J1856.5-3754 has been imaged multiple times by \textit{HST}, making it
an ideal candidate for detailed analysis. Its position, proper motion
and parallax have been measured on several occasions
\citep{2001ApJ...549..433W,2002ApJ...576L.145W,2010ApJ...724..669W}. Using 
these measurements, we take $D_L=123\pm 15$ pc, and parallax
$\pi=8.2\pm0.2$ mas. We then calculate the Einstein angle (with
$M_L=1.4\, M_\odot$) to be $\theta_E=9.02\pm 0.61$ mas.

\begin{figure}
  \centering
  \includegraphics[width=0.48\textwidth]{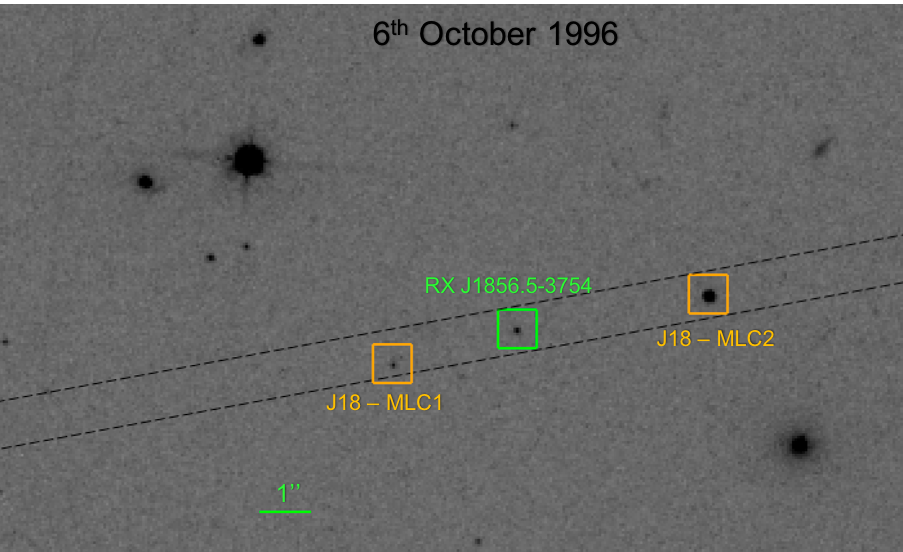}
  \includegraphics[width=0.48\textwidth]{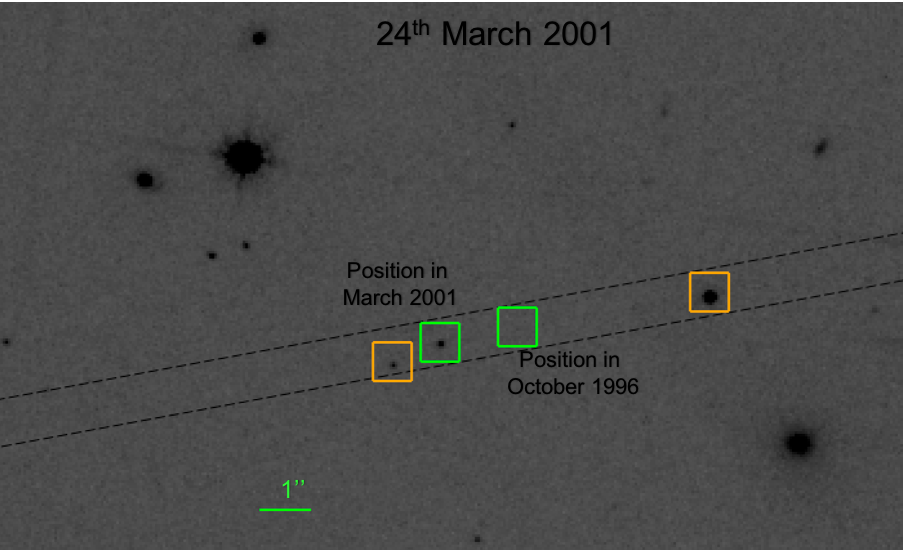}
  \caption[Steady flow]{Two images of RX J1856.5-3754 taken by \textit{HST's}
    WFPC2 camera in the F606W filter at different epochs. The two
    dashed black lines run parallel to RX J1856.5-3754's projected
    path. {\sl Top:} 6$^{th}$ October 1996, {\sl Bottom:} 24$^{th}$
    March 2001.}
  \label{J1856}
\end{figure}

We retrieved two \textit{HST} images of RX J1856.5-3754 from the HLA
(epochs: 1996, 2001). They were taken by the PC1 chip of the WFPC2
camera with the F606W filter (wide V band). The images were uploaded
to DS9 and the neutron star's path overlaid (Fig. \ref{J1856}). It
became apparent that two lensing events could
have occurred in the past. For convenience 
we will refer to the two relevant background sources 
as J18-MLC1\footnote{MLC - Microlensing Candidate} and J18-MLC2; 
these are stars $\#$115 and $\#$23 in
\citet{2001ApJ...549..433W}. J18-MLC1 does not appear in any source
catalogue. Note that this is an example (like the one shown in
Fig. \ref{GRO}) in
which non-catalogued sources on the images can only be identified through
visual examination. 

RX J1856.5-3754's proper motion is noticeable when comparing the two
images. In contrast, there is no observable change in position for J18-MLC1 between
the two images, suggesting a low proper motion, in
agreement with 
\citet{2002ApJ...571..447K}. J18-MLC2's motion is much more prominent
and has been measured by \citet{2010ApJ...724..669W}. The motion due to of
RX J1856.5-3754's parallax would be a third of the size of a pixel in this
image. 

We computed the distances and times of closest
approach. We take the starting positions of the lens and both sources from the
1996 image. By assuming that the uncertainties in position are the same for
all the objects in the image, we have considered the relative positional errors
to be negligible.
We computed the closest distance of approach, $\delta\phi_c$, of RX J1856.5-3754 to J18-MLC1 to be
$\delta\phi_c = 194.94\pm 2$ mas, occurring on 25th April 2004. This
estimate does not agree with the prediction made by
\citet{2001astro.ph..7443P}, which estimated a closest approach to
occur in June 2003, at a separation of $\sim 300$ mas. However
\citet{2002ApJ...571..447K} revised this prediction and claimed an
approach of distance $200$ mas in April 2004. This result agrees with
our estimated prediction. We find the approach to be closer when
taking J18-MLC1's  proper motion into account: $\delta\phi_c = 167.78$
mas; the new time of approach is three days later. An
approach of this distance would shift J18-MLC1's centroid position by
$\sim0.48\pm 0.07$ mas, which is potentially detectable. 
Eight images were taken in the F475W filter with \textit{HST's}
ACS/HRC camera during 2002$\--$2004. It is difficult to locate the source
in the images but the observation taken in May 2004 appears to show RX
J1856.5-3754 very close to J18-MLC1.

\begin{figure}
\centering
\includegraphics[width=0.49\textwidth]{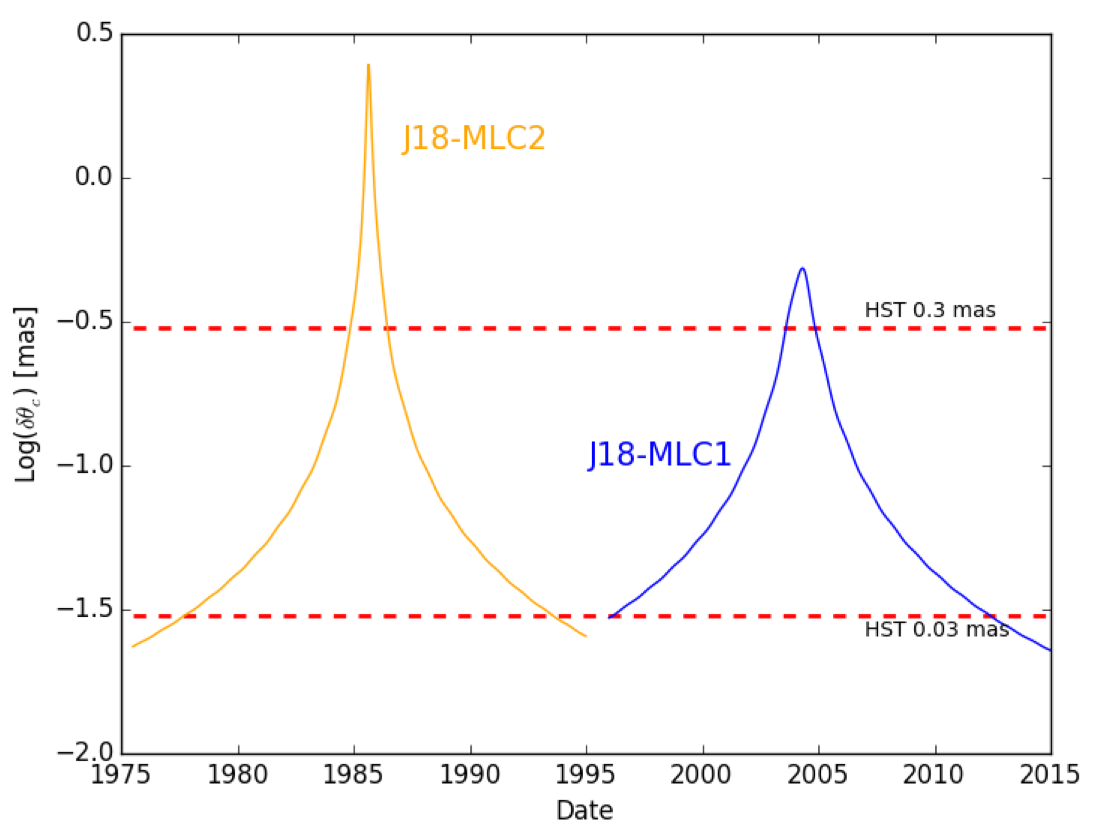}
\caption{Predicted microlensing centroid shifts of the background
  sources J18-MLC1 and J18-MLC2, from a close passage of RX
  J1856.5-3754. The two red cut-off lines at $0.3$ and $0.03$ mas
  are the limits of possible detectability suggested by 
  \citet{2011PASP..123..622B} and \citet{Riess:2014uga} respectively. 
  The close approach distances would cause detectable centroid shifts 
  in both sources.} 
  \label{J1856-shifters}
\end{figure}

We also find that J18-MLC2 was lensed in August 1985. 
The closest approach distance, $\delta\phi_c=33.23 \pm 4$, mas would
have produced a detectable astrometric shift of $2.45\pm0.7$ mas,
and a small photometric effect ($\sim 1$ per cent) as well.

We note that the durations of detectable events are determined by the
detectability limit. We consider that the start (end) of an event occurs when the size of the 
centroid shift is greater (less) than the specified detectability limit. Fig. 
\ref{J1856-shifters} shows that for a $0.3$ mas cut-off, the predicted events 
would be detectable for $\sim~1$ year. If smaller shifts of $0.03$ mas were 
detectable, then approaches as close as these would produce events 
observable for time intervals on the order of decades.

\subsection{PM I12506+4110E}

The white dwarf PM I12506+4110E is one of 62 white dwarfs for which
we had HSC source lists and identified stars  
lying within $10$ arcsec of its path. Because our simulation registered
a close approach distance of $60$ mas to a background HSC source, we
analysed the corresponding \textit{HST} images. The white dwarf was
imaged during 2005 in two filters: the F555W and F435W filters on
the ACS/WFC camera. 
In the image we found a second star near the white dwarf's path.
Neither of the stars that will be approached by the white dwarf are listed 
in catalogues covered by the VizieR service. 

\begin{figure}
\centering
\includegraphics[width=0.48\textwidth]{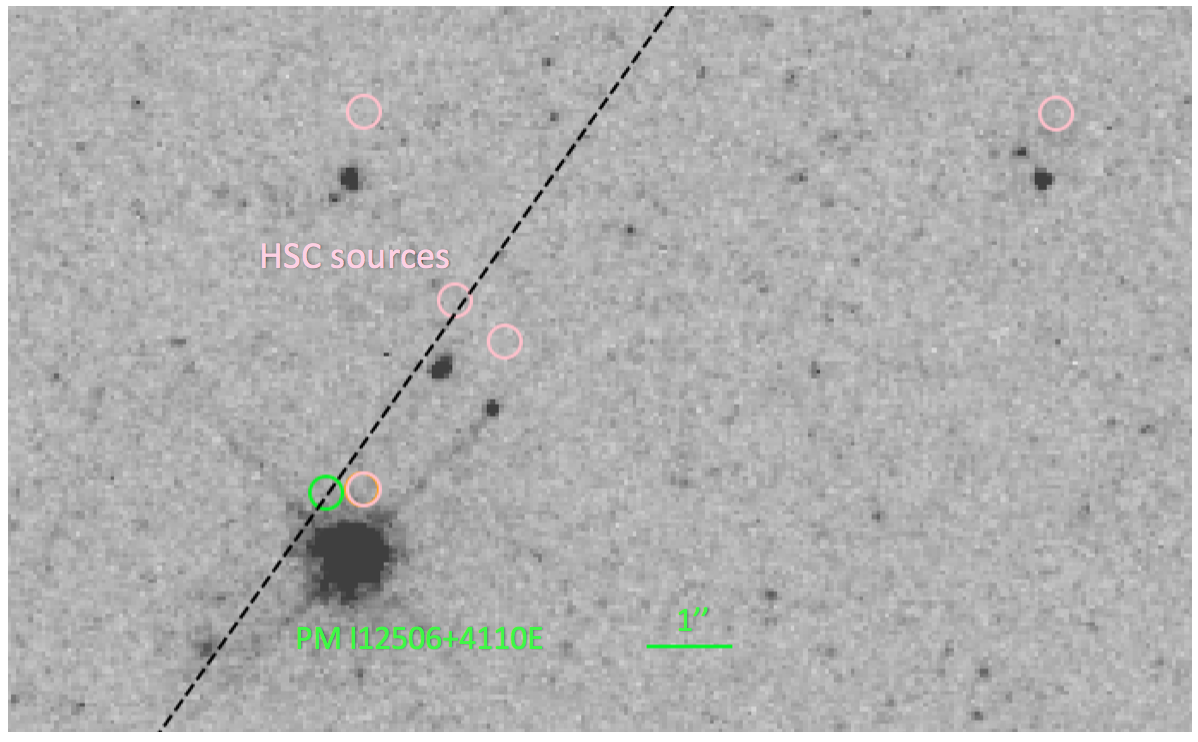}
\includegraphics[width=0.48\textwidth]{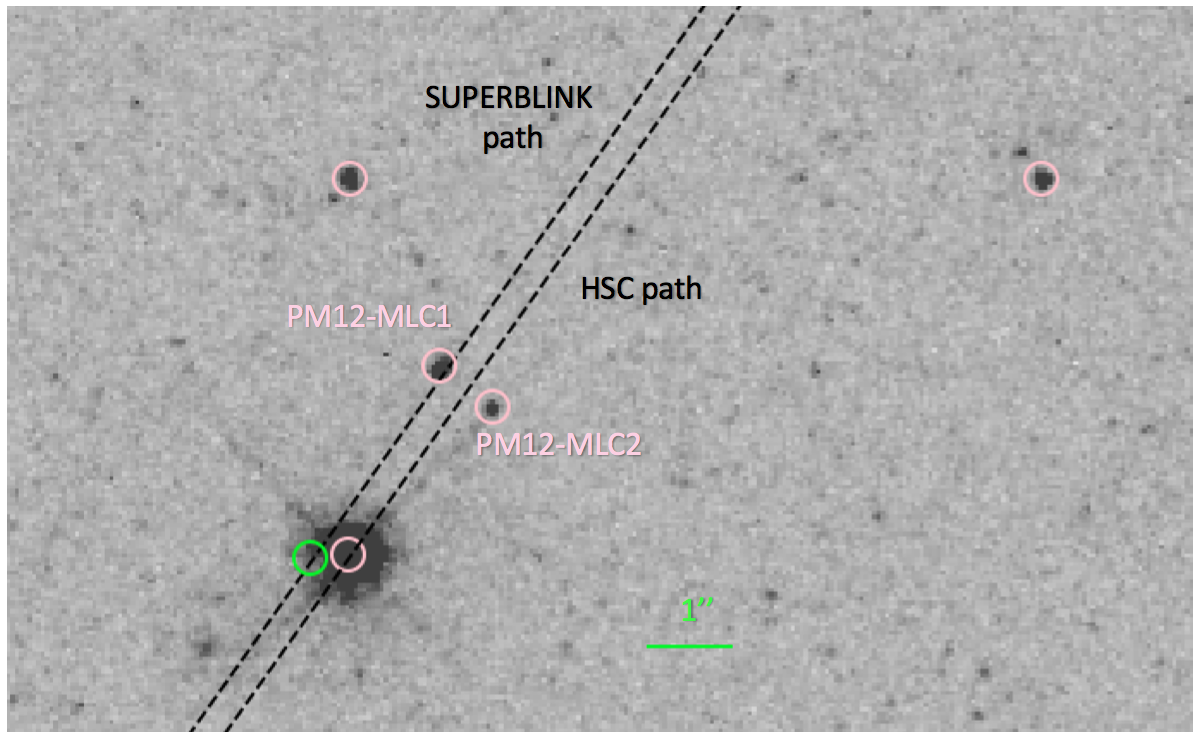}
\caption{F555W \textit{HST} image from the ACS/WFC camera taken in 2005 of
  the white dwarf PM I12506+4110E. The lens's path passes directly
  between two background sources: PM12-MLC1 and PM12-MLC2. {\sl Top:}
  shows the image with the HSC sources positions (pink) and position
  of the lens from SUPERBLINK (green) overlaid. The black hashed line
  shows the projected path of the lens originating for the SUPERBLINK
  position. {\sl Bottom:} The source and lens positions have been
  calibrated by addition of an average shift. There are now two black
  lines representing the projected path, one is for the lens's
  position from SUPERBLINK, the other from the HSC position. We use
  the HSC position to calculate the close approaches.}
\label{PM_I12PM_I12}
\end{figure}

Fig. \ref{PM_I12PM_I12} shows the same image in both panels. The top
panel shows the overlaid source positions from HSC (pink): one of
these is PM I12506+4110E. The green circle is the position
of the lens as computed from information in the SUPERBLINK catalogue. 
We made astrometric corrections to the image to produce
the bottom panel of the Figure. The average shift between the HSC
quoted positions and the positions on the image was calculated. 
Because we focus on this small area of the image,
we used only the sources near the lens path to calculate the average
shift. 
We
applied the average shift to the HSC source positions.
Even after this process was implemented, the \wdf \ position
derived from the SUPERBLINK catalogue
is offset compared to the position of the white dwarf in the
image. The paths based on both the SUPERBLINK and HSC positions have
been plotted in the lower panel. The discrepancy is likely due to the
position uncertainty from the SUPERBLINK catalogue. We use the path
projected from the HSC position to make the event prediction. 

We refer to the two background sources as PM12-MLC1 and PM12-MLC2. Using
the proper motion provided by SUPERBLINK and the position of the lens from
the HSC, we project the path of PM I12506+4110E. This path passes directly
between the two sources. We predict approaches of $402.71$ mas to
PM12-MLC1 and $346.23$ mas to PM12-MLC2 occurring in 2042 and 2044
respectively.  Neither would produce a centroid shift $>0.3$ mas. With
maximum uncertainties of $7$ mas yr$^{-1}$ on the proper motions
\citep{2005AJ....129.1483L}, over time, the path may diverge from the one 
we have projected. All possible paths of the lens will exist inside
a cone with sides relating to the paths calculated with these maximum
uncertainties. A path along one of these sides will come within $u
\sim 9~\theta_E$ of PM12-MLC1, whereas a path along the other side
will approach PM12-MLC2 at a distance of $u \sim 6~\theta_E$. Both of
these approaches would produce centroid shifts of less than $1$ mas
but greater than $0.3$ mas. This is an interesting case as uncertainties
in the path provide a chance for PM I12506+4110E to pass close to one
of the sources. Further imaging of this white dwarf will allow for
improvements to the accuracy of our prediction. More observations will
also be required to check for any proper motion in the two background
sources.

\section{Conclusions}

We have demonstrated that the prediction of lensing events is 
a practical idea. The existence of deep catalogues of background
objects is making work along these lines ever more practical. We applied these ideas to stellar remnants because of the
importance of measuring their masses, a task to which gravitational 
microlensing is well suited.
In addition, stellar remnants tend to be dim, making 
it easier to discern the changes wrought by lensing in the positions
and brightness of background stars.

We find that the detection of lensing events due to white dwarfs
can certainly be observed during the next decade by both \textit{Gaia} and \textit{HST} \citep{2017AAS...23031513S}.
Photometric events will occur, but to detect them will require 
observations of the positions
of hundreds to thousands of far-flung white dwarfs.
As we learn the positions, distances to, and proper motions of larger
numbers of white dwarfs through the completion of surveys such as \textit{Gaia}
and through ongoing and new wide-field surveys, the situation will continue to
improve.

For neutron stars and black holes, the regular and systematic prediction of
lensing events will be more challenging. The known populations are too
small, and we do not have measured proper motions and/or distances even
for the majority of the known systems. 
As new discoveries increase the known populations and as
distance measurements and determinations
of proper motions are made, the prospects for prediction will improve.
Despite the low rate, we have found that at least one neutron
star came close to
producing detectable events in the recent past. This indicates the 
value of high-resolution imaging for compact objects that happen to be 
moving in front of dense stellar fields.

Within the next several years, the measurements of white dwarf masses 
seems poised to become a possible enterprise that will open a new window to the
study of compact objects \citep{2017AAS...23031513S}.  
The discovery of planets orbiting  compact objects will be an additional
benefit of these studies.

\section*{Acknowledgements}

R.D. thanks the Smithsonian Institution's Scholarly Studies program for support.
J.U. thanks the the Harvard-Smithsonian Center for Astrophysics for its hospitality during the early stages of this work, and the IAU for support. This work has made use of data from the European Space Agency (ESA)
mission {\it Gaia} (\url{http://www.cosmos.esa.int/gaia}), processed by
the {\it Gaia} Data Processing and Analysis Consortium (DPAC,
\url{http://www.cosmos.esa.int/web/gaia/dpac/consortium}). Funding
for the DPAC has been provided by national institutions, in particular
the institutions participating in the {\it Gaia} Multilateral Agreement. This work has used observations made with the NASA/ESA Hubble Space Telescope, and obtained from the Hubble Legacy Archive, which is a collaboration between the Space Telescope Science Institute (STScI/NASA), the Space Telescope European Coordinating Facility (ST-ECF/ESA) and the Canadian Astronomy Data Centre (CADC/NRC/CSA). This research has made use of the VizieR catalogue access tool, CDS, Strasbourg, France. The original description of the VizieR service was published in A\&AS 143, 23.




\bibliographystyle{mnras}
\bibliography{refs.bib} 


  


\bsp	
\label{lastpage}
\end{document}